\documentclass[prodmode,acmtist]{acmsmall} % Aptara syntax

\usepackage[ruled]{algorithm2e}

\SetAlFnt{\small}
\SetAlCapFnt{\small}
\SetAlCapNameFnt{\small}
\SetAlCapHSkip{0pt}
\IncMargin{-\parindent}

\acmVolume{0}
\acmNumber{0}
\acmArticle{0}
\acmYear{2016}
\acmMonth{0}

\usepackage{floatrow}
\usepackage{subfigure}

\usepackage{multirow}
\usepackage{array}

\newcommand{\squishlist}{
\begin{list}
	{$\bullet$} { \setlength{
	\itemsep}{0pt} \setlength{\parsep}{3pt} \setlength{\topsep}{3pt} \setlength{
	\partopsep}{0pt} \setlength{\leftmargin}{1.5em} \setlength{\labelwidth}{1em} \setlength{\labelsep}{0.5em} } }
	
	\newcommand{\squishlisttwo}{
	\begin{list}
		{$\bullet$} { \setlength{
		\itemsep}{0pt} \setlength{\parsep}{0pt} \setlength{\topsep}{0pt} \setlength{
		\partopsep}{0pt} \setlength{\leftmargin}{2em} \setlength{\labelwidth}{1.5em} \setlength{\labelsep}{0.5em} } }
		
		\newcommand{\squishend}{
	\end{list}
	}

\doi{0000001.0000001}

\issn{1234-56789}

\begin{document}

% Page heads
\markboth{T. Tran et al.}{How to Succeed in Crowdfunding: a Long-Term Study in Kickstarter}

% Title portion
\title{How to Succeed in Crowdfunding: a Long-Term Study in Kickstarter}
\author{
THANH TRAN
\affil{Utah State University}
MADHAVI R. DONTHAM
\affil{Utah State University}
JINWOOK CHUNG
\affil{Utah State University}
KYUMIN LEE
\affil{Utah State University}
}
% NOTE! Affiliations placed here should be for the institution where the
%       BULK of the research was done. If the author has gone to a new
%       institution, before publication, the (above) affiliation should NOT be changed.
%       The authors 'current' address may be given in the "Author's addresses:" block (below).
%       So for example, Mr. Abdelzaher, the bulk of the research was done at UIUC, and he is
%       currently affiliated with NASA.

\begin{abstract}
Crowdfunding platforms have become important sites where people can create projects to seek funds toward turning their ideas into products, and back someone else's projects. As news media have reported successfully funded projects (e.g., Pebble Time, Coolest Cooler), more people have joined crowdfunding platforms and launched projects. But in spite of rapid growth of the number of users and projects, a project success rate at large has been decreasing because of launching projects without enough preparation and experience. Little is known about what reactions project creators made (e.g., giving up or making the failed projects better) when projects failed, and what types of successful projects we can find. To solve these problems, in this manuscript we (i) collect the largest datasets from Kickstarter, consisting of all project profiles, corresponding user profiles, projects' temporal data and users' social media information; (ii) analyze characteristics of successful projects, behaviors of users and understand dynamics of the crowdfunding platform; (iii) propose novel statistical approaches to predict whether a project will be successful and a range of expected pledged money of the project; (iv) develop predictive models and evaluate performance of the models; (v) analyze what reactions project creators had when project failed, and if they did not give up, how they made the failed projects successful; and (vi) cluster successful projects by their evolutional patterns of pledged money toward understanding what efforts project creators should make in order to get more pledged money. Our experimental results show that the predictive models can effectively predict project success and a range of expected pledged money.
\end{abstract}




%\begin{CCSXML}
%<ccs2012>
%  <concept_id>10010520.10010553.10010562</concept_id>
%  <concept_desc>Computer systems organization~Embedded systems</concept_desc>
%  <concept_significance>500</concept_significance>
%  <concept_id>10010520.10010575.10010755</concept_id>
%  <concept_desc>Computer systems organization~Redundancy</concept_desc>
%  <concept_significance>300</concept_significance>
%  <concept_id>10010520.10010553.10010554</concept_id>
%  <concept_desc>Computer systems organization~Robotics</concept_desc>
%  <concept_significance>100</concept_significance>
%  <concept_id>10003033.10003083.10003095</concept_id>
%  <concept_desc>Networks~Network reliability</concept_desc>
%  <concept_significance>100</concept_significance>
%</ccs2012>
%\end{CCSXML}

%\ccsdesc[500]{Computer systems organization~Embedded systems}


% We no longer use \terms command
%\terms{Design, Algorithms, Performance}

\keywords{Crowdfunding, kickstarter, twitter, project success, fundraising amount, clustering projects}

\acmformat{Thanh Tran, Madhavi R. Dontham, Jinwook Chung, Kyumin Lee, 2016. How to Succeed in Crowdfunding: a Long-Term Study in Kickstarter.}
% At a minimum you need to supply the author names, year and a title.
% IMPORTANT:
% Full first names whenever they are known, surname last, followed by a period.
% In the case of two authors, 'and' is placed between them.
% In the case of three or more authors, the serial comma is used, that is, all author names
% except the last one but including the penultimate author's name are followed by a comma,
% and then 'and' is placed before the final author's name.
% If only first and middle initials are known, then each initial
% is followed by a period and they are separated by a space.
% The remaining information (journal title, volume, article number, date, etc.) is 'auto-generated'.

\begin{bottomstuff}
An early version of this manuscript appeared in the 2015 ACM Proceedings of the Hypertext \& Social Media conference \cite{Chung:2015}.

Author's addresses: T. Tran, M. R. Dontham, J. Chung, and K. Lee, Department of Computer Science, Utah State University, Logan, UT 84341; email: thanh.tran@aggiemail.usu.edu, madhavidontham@aggiemail.usu.edu, jinwookchung.jin@gmail.com, kyumin.lee@usu.edu.
\end{bottomstuff}

\maketitle

\section{Introduction}
Crowdfunding platforms have successfully connected millions of individual crowdfunding backers to a variety of new ventures and projects, and these backers have spent over a billion dollars on these ventures and projects \cite{Gerber:2013}. From reward-based crowdfunding platforms like Kickstarter, Indiegogo, and RocketHub, to donation-based crowdfunding platforms like GoFundMe and GiveForwad, to equity-based crowdfunding platforms like CrowdCube, EarlyShares and Seedrs - these platforms have shown the effectiveness of funding projects from millions of individual users. The US Congress has encouraged crowdfunding as a source of capital for new ventures via the JOBS Act \cite{jumpstart}.

An example of successfully funded projects is E-paper watch project. The E-paper watch project for smartphones on a crowdfunding platform was created by Pebble Technology corporation on April 2012 in Kickstarter, expecting \$100,000 investment. Surprisingly, in 2 hours right after launching the project, pledged money was already exceeding \$100,000. In the end of the project period (about 5 weeks), the company was able to get investment over 10 million dollars \cite{pebble}. This example shows the power of collective investment and a crowdfunding platform, and a new way to raise funding from the crowds.

Even though the number of projects and amount of pledged funds on crowdfunding platforms has dramatically grown in the past few years, success rate of projects at large has been decreasing. Besides, little is known about dynamics of crowdfunding platforms and strategies to make a project successful. To fill the gap, in this manuscript we are interested to (i) analyze Kickstarter, the most popular crowdfunding platform and the 524th most popular site as of March 2016 \cite{alexa};(ii) propose statistical approaches to predict not only whether a project will be successful, but also how much a project will get invested; (iii) understand What reactions project creators made when their projects failed; and (iv) find successful project groups, and understand how they are different. Kickstarter has an All-or-Nothing policy. If a project reaches pledged money lower than its goal, its creator will receive nothing. Predicting a range of expected pledged money is an important research problem.

Specifically, we analyze behaviors of users on Kickstarter by answering following research questions: Are users only interested in creating and launching their own projects? or Do they support other projects? Has the number of newly joined users increased over time? Have experienced users achieved a higher project success rate? Then, we analyze characteristics of projects by answering following research questions: How many projects have been created over time? What percent of project has been successfully funded? Can we observe distinguishing characteristics between successful projects and failed projects? Based on the analysis and study, we answer following research questions: Can we build predictive models which can predict not only whether a project will be successful, but also a range of expected pledged money of the project? By adding a project's temporal data (e.g., daily pledged money and daily increased number of backers) and a project creator's social media information, can we even improve performance of the predicative models further? Other interesting questions are: What reactions did project creators make when project failed? If they re-launched the failed projects with some improvements, what efforts did they make for success of the projects? By clustering successful projects, can we understand how we can even further increase pledged money based on understanding properties of more successful projects with higher pledged moneys?

Toward answering these questions, we make the following contributions in this manuscript:
\begin{itemize}
\item We collected the largest datasets, consisting of all Kickstarter project pages, user pages, each project's temporal data and each user's Twitter account information, and then conducted comprehensive analysis to understand behaviors of Kickstarter users and characteristics of projects.
\item Based on the analysis, we proposed and extracted four types of features toward developing project success predictors and pledged money range predictors. To our knowledge, this is the first work to study how to predict a range of expected pledged money of a project.
\item We developed predictive models and thoroughly evaluated performance of these models. Our experimental results show that these models can effectively predict whether a project will be successful and a range of expected pledged money.
\item We analyzed what reactions project creators had when project failed. If they re-launched the failed projects with some improvements and made them successful, what efforts they would make.
\item Finally, we clustered successful projects toward understanding how these clusters are different and revealing what strategy projects creators should use to increase pledged money.
\end{itemize}

\section{Related Work}
In this section we summarize crowdfunding research work in four categories: (i) analysis of crowdfunding platforms; (ii) analysis of crowdfunding activities and backers on social media sites; (iii) project success prediction; and (iv) classification of backers or projects.

Researchers have analyzed crowdfunding platforms \cite{Belleflamme:2012,Gerber:2013,gerber2012crowdfunding,hui2014understanding}. For example, Kuppuswamy and Bayus \cite{Kuppuswamy} examined the backer dynamics over the project funding cycle. Mollick \cite{mollick:2014} studied the dynamics of crowdfunding, and found that personal networks and underlying project quality were associated with the success of crowdfunding efforts. Xu et al. \cite{Xu:2014} analyzed the content and usage patterns of a large corpus of project updates on Kickstarter. Joenssen et al. \cite{joenssen2014link} found that timing and communication (by posting updates) were key factors to make project successful. Joenssen and M{\"u}llerleile \cite{joenssen2016limitless} analyzed 42,996 Indiegogo projects, and found that scarcity management was problematic at
best and reduced the chances of projects to successfully achieve their target funding. Althoff and Leskovec \cite{althoff2015donor} presented various factors impacting investor's retention, and identified various types of investors. The researchers found that investors are more likely to return if they had a positive interaction with the receiver of the funds.

In another research direction, researchers have studied social media activities during running project campaigns on crowdfunding platforms. Lu et al. \cite{lu:2014} studied how fundraising activities and promotional activities on social media simultaneously evolved over time, and how the promotion campaigns influenced the final outcomes. \citeN{Rakesh:2015} used a promoter network on Twitter to show the success of projects depended on the connectivity between the promoters. They developed backer recommender which recommends a set of backers to Kickstarter projects. Lu et al. \cite{lu2014identifying} analyzed the hidden connections between the fundraising results of projects on crowdfunding websites and the corresponding promotion campaigns in social media. An et al. \cite{an2014recommending} proposed different ways of recommending investors by using hypothesis-driven analyses. Naroditskiy et al. \cite{naroditskiy2014referral} investigated whether viral marketing with incentive mechanisms would increase the marketing and found that providing high level of incentives resulted in a statistically significant increase.

Predicting the success of a project is one of important research problems, so researchers have studied how to predict whether a project will be successful or not. Greenberg et al. \cite{Greenberg:2013} collected 13,000 project pages on Kickstarter and extracted 13 features from each project page. They developed classifiers to predict project success. Their approach achieved 68\% accuracy. Etter et al. \cite{Etter:2013} extracted pledged money based time series features, and project and backer graph features from 16,000 Kickstarter projects. Then, they measured how prediction rate has been changed over time. Mitra et al. \cite{Mitra:2014} focused on text features of project pages to predict project success. They extracted phrases and some meta features from 45,810 project pages, and then showed that using phrases features reduced prediction error rates. Xu et al. \cite{Xu:2014} investigated how updates influence the outcome of a project and showed the type of updates that had a positive impact in every stage of a project. Solomon et al. \cite{solomon2015don} found that making an early donation was usually a better strategy for donors because the amount of donations made early in a project's campaign was often the only difference between that project being funded or not.

Other researchers have classified backers and projects to various types. Kuppuswamy and Bayus \cite{kuppuswamy2015crowdfunding} classified backers into three categories -- immediate backers, delayed backers and serial backers. Hemer \cite{hemer2011snapshot} classified crowdfunding projects into for-profit or not-for-profit projects. Haas et al. \cite{haas2014empirical} also classified projects into hedonistic or altruistic projects using a clustering algorithm from a business standpoint.

Compared with the previous research work, we collected the largest datasets consisting of all Kickstarter project pages, corresponding user pages, each project's temporal data and each user's social media profiles, and conducted comprehensive analysis of users and projects. Then, we proposed and extracted comprehensive feature sets (e.g., project features, user features, temporal features and Twitter features) toward building project success predictors and pledged money range predictors. To our knowledge, we are the first to study how to predict a range of expected pledged money of a project. Since the success of a project depends on a project goal and the amount of actually pledged money, studying the prediction is very important. In addition, we analyzed when project failed what efforts project creators made for success of the projects. Finally, by using a Gaussian mixture model based clustering algorithm, we clustered successful projects to understand how these clusters were different and how project creators increase pledged money. Our research will complement the existing research base.

\section{Datasets}
\label{sec:dataset}

To analyze projects and users on crowdfunding platforms, and understand whether adding social media information would improve project success prediction and pledged money prediction rates, what kind of successful project groups we could find, first we collected data from Kickstarter, the most popular crowdfunding platform, and Twitter, one of the most popular social media sites. The following subsections present our data collection strategy and datasets.

\subsection{Kickstarter Dataset}
%*Static Data*
Kickstarter is a popular crowdfunding platform where users create and back projects. As of March 2016, it is the 524th most visited site in the world according to Alexa \cite{alexa}.

\smallskip
\noindent\textbf{Static Data.} Our Kickstarter data collection goal was to collect all Kickstarter pages and corresponding user pages, but Kickstarter site only shows currently active projects and some of the most funded projects. Fortunately, Kicktraq site\footnote{\url{http://www.kicktraq.com/archive/}} has archived all project page URLs of Kickstarter. Given a Kicktraq project URL\footnote{\url{http://www.kicktraq.com/projects/fpa/launch-the-first-person-arts-podcast/}}, by replacing Kicktraq hostname (i.e, \url{www.kicktraq.com}) of the project URL with Kickstarter hostname (i.e., \url{www.kickstarter.com}), we were able to obtain the Kickstarter project page URL\footnote{\url{https://www.kickstarter.com/projects/fpa/launch-the-first-person-arts-podcast/}}.

Specifically, our data collection approach was to collect all project pages on Kicktraq, extract each project URL, and replace its hostname with Kickstarter hostname. Then we collected each Kickstarter project page and corresponding user page. Note that even though Kickstarter do not reveal an old project page (i.e., a project's campaign duration was ended), if we know the project URL, we can still access the project page on Kickstarter. %This is the first study to collect and analyze all project pages on Kickstarter.

Finally, we collected 168,851 project pages which were created between 2009 and September 2014. Note that Kickstarter site was launched in 2009. A project page consists of a project duration, funding goal, project description, rewards description and so on. We also collected corresponding 146,721 distinct user pages each of which consists of bio, account longevity, location information, the number of backed projects, the number of created projects, and so on. Among 168,851 project pages, we filtered 17,243 projects which have been either canceled or suspended, or in which the project creator's account has been canceled or suspended. Among 146,721 user pages, we filtered corresponding 14,435 user pages. Finally, 151,608 project pages and 132,286 user pages presented in Table~\ref{table:dataset}, have been used in the rest of this manuscript.

\smallskip
\noindent\textbf{Temporal Data.} To analyze and understand how much each project has been pledged/invested daily and how many backers each project has attracted daily, whether incorporating these temporal data (i.e., daily pledged money and daily increased number of backers during a project duration) can improve project success prediction and expected pledged money prediction rates, we collected temporal data of 74,053 projects which were created between March 2013 and August 2014 and were ended by September 2014.

\begin{table}
	\tbl{Datasets. \label{table:dataset}}
	{
		\centering
		\small
		\begin{tabular}{|l|c|c|ll}
			\hline
			Kickstarter projects  & 151,608 \\ \hline
			Kickstarter users     & 132,286 \\ \hline
			\hline
			Kickstarter projects with temporal data & 74,053 \\ \hline
			Kickstarter projects with Twitter user profiles & 21,028 \\ \hline
		\end{tabular}
	}
	%\vspace{-5pt}
\end{table}

\subsection{Twitter Dataset} What if we add social media information of a project creator to build predictive models? Can a project creator's social media information improve project success and expected pledged money prediction rates? Can we link a project creator's account on Kickstarter to Twitter? To answer these questions, we checked project creators' Kickstarter profiles. Interestingly 19,138 users (13.4\% of all users in our dataset), who created 22,408 projects, linked their Twitter user profile pages (i.e., URLs) to their Kickstarter user profile pages. To use these users' Twitter account information in experiments, we collected their Twitter account information. Specifically, we extracted a Twitter user profile URL from each Kickstarter user profile, and then collected the user's Twitter profile information consisting of the basic profile information (e.g., a number of tweets, a number of following and a number of followers) and tweets posted during a project period. In a step of the Twitter user profile collection, we noticed that some of Twitter accounts had been either suspended or deleted. By filtering these accounts, finally, we collected 17,908 Twitter user profiles and tweets, and then combined these Twitter information with 21,028 Kickstarter project pages created by the 17,908 users.

\section{Analyzing Kickstarter Users and Projects}
In the previous section, we presented our data collection strategy and datasets. Now we turn to analyze Kickstarter users and projects.

\begin{figure}[!ht]
    \centering
    \begin{floatrow}
      \ffigbox[\FBwidth]
      {
      		\caption{Number of newly joined Kickstarter users in each month.}
      		\label{fig:change-date}
      }
      {
	  		\includegraphics[width=0.465\textwidth]{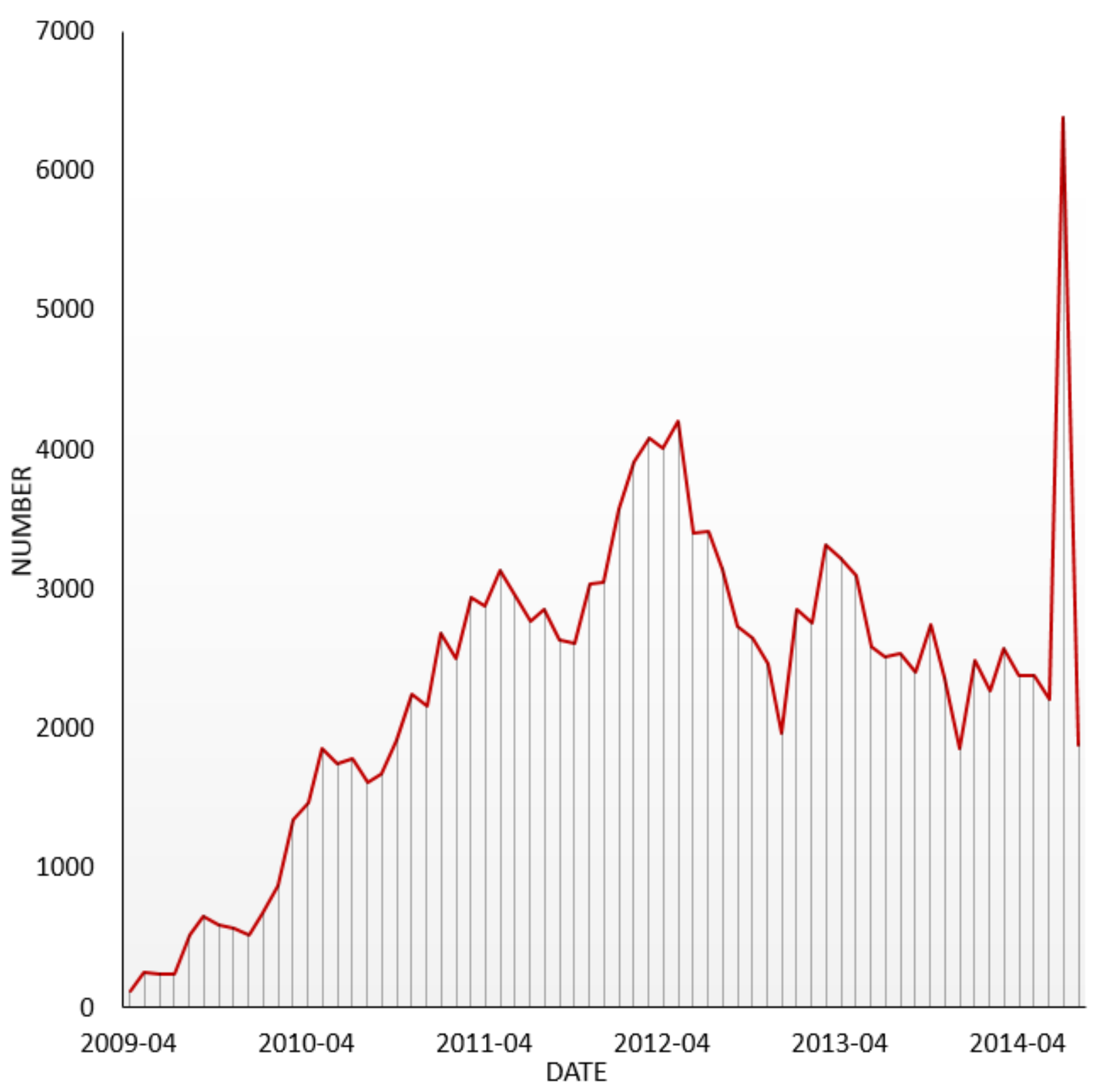}
      }
      \ffigbox[\FBwidth]
      	{	
      		\caption{CDFs of intervals between user joined date and project creation date (Days).}
      		\label{fig:interval}
      	}
      	{
        		\includegraphics[width=0.465\textwidth]{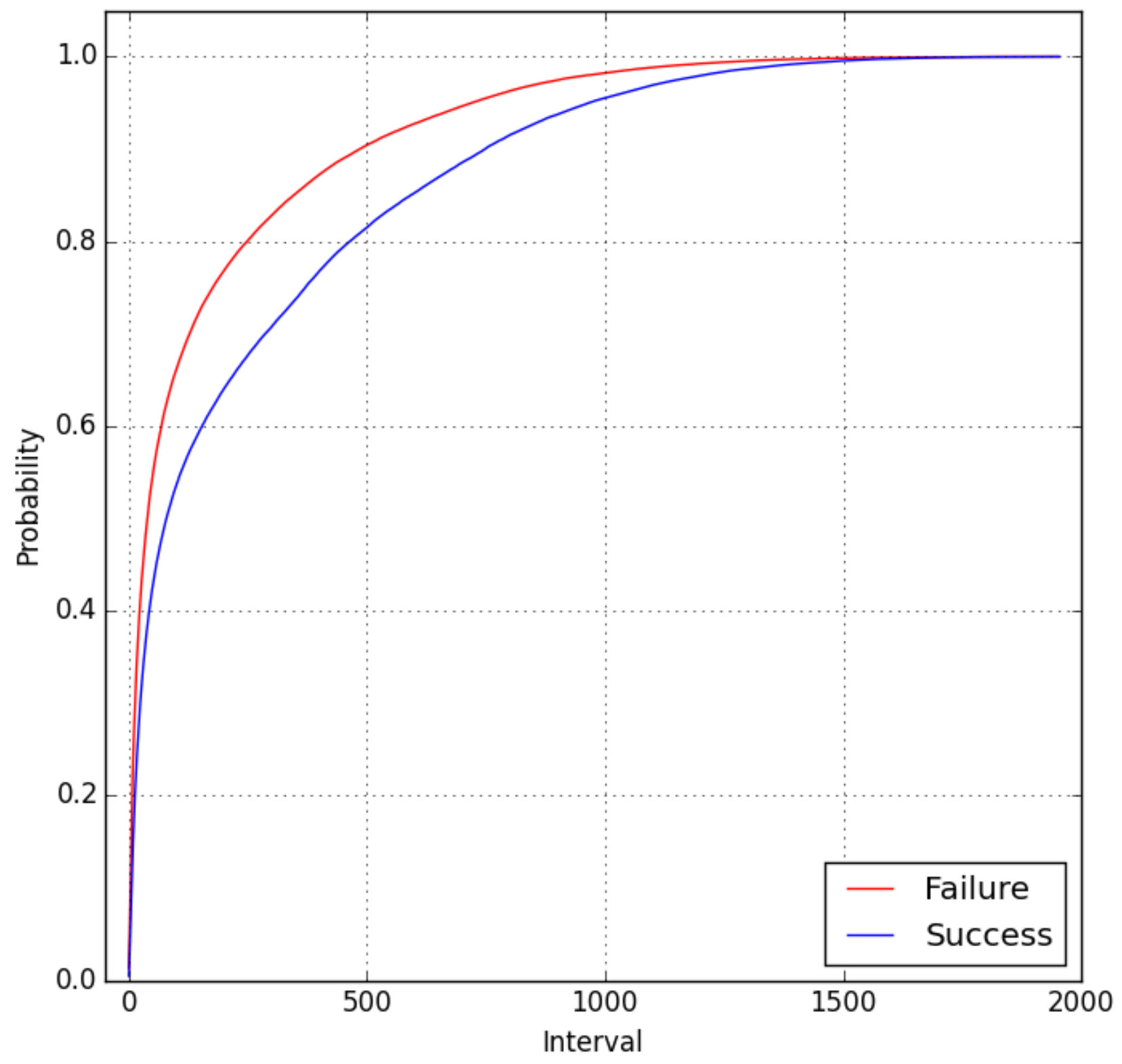}
      	}	
    \end{floatrow}
\end{figure}

%[subsub-section title]
\subsection{Analysis of Users}
Given 132,286 user profiles, we are interested in answering following research questions: Has the number of newly joined users increased over time? Are users only interested in creating and launching their own projects? or Do they support other projects? Do experienced users have a higher probability to make a project successful?

First of all, we analyze how many new users joined Kickstarter over time. Figure~\ref{fig:change-date} shows the number of newly joined Kickstarter users per month. Overall, the number of newly joined users per month has linearly increased until May 2012, and then has been decreased until June 2014 with some fluctuation. In July 2014, there was a huge spike. Note that we tried to understand why there was a huge spike in July 2014 by checking news articles, but we were not able to find a concrete reason. Interesting observation is that the number of newly joined users was the lowest during winter season, especially, December in each year. We conjecture that since November and December contains several holidays, people may delay to join Kickstarter.

%After joining Kickstarter, when do they create their first project?
Next, we present general statistics of users in Table~\ref{fig:gs-creator-page}. The user statistics show that average number of backed projects and created projects are 3.48 and 1.19, respectively. It means that users backed larger number of projects and created less number of their own projects. Each user linked 1.75 websites on average into her profile so that she can get trust from potential investors. Examples of websites are company sites and user profile pages in social networking sites such as Twitter and YouTube. 13.4\% Kickstarter users linked their Twitter pages, and 6.89\% Kickstarter users linked their YouTube pages.

\begin{table}[t]%[h]
\tbl{Statistics of Kickstarter users. \label{fig:gs-creator-page}}
{
\centering
\small
\begin{tabular}{lc}
                                & \textbf{Total} \\ \hline
Total number of users        & 132,286        \\
Number of backed projects per user               & 3.48           \\
Number of created projects per user              & 1.19           \\
%Number of comments creator left & 6.67           \\
Number of websites per user  & 1.75           \\
Twitter connected          & 13.4\% users          \\
YouTube connected          & 6.89\% users         \\ \hline
\end{tabular}
}
%\vspace{-5pt}
\end{table}

\begin{table}[t]%[h]
\tbl{Two groups of users: all-time (AT) creators and active users \label{table:groups}}
{
\centering
\small
\begin{tabular}{rrrr}
\multicolumn{1}{l}{} & \multicolumn{1}{c}{\textbf{Number}} & \multicolumn{1}{c}{\textbf{Avg. backed}} & \multicolumn{1}{c}{\textbf{Avg. created}} \\ \hline
%AT-backers           & 17                                  & 3.1176                                   & N/A                                       \\
%AT creators        & 66,262                              & N/A                                      & 1.12                                    \\
%Active users         & 76,628                             & 6.49                                   & 1.25                                    \\
AT creators			& 60,967                              	& N/A                                 	& 1.12                                    \\
Active users		& 71,319                             	& 6.45                                  & 1.25                                    \\
%Inactive users       & 22                                  & N/A                                      & N/A                                       \\ \hline
\end{tabular}
}
%\vspace{-5pt}
\end{table}

%\smallskip
%\noindent\textbf{All-time creators and active users.}
Next, we categorized Kickstarter users based on their project backing and creating activities. We found two groups of users: (i) all-time creator (AT creator), who only created projects and did not back other projects; and (ii) active user, who not only created her own projects but also backed other projects. As shown in Table~\ref{table:groups}, there are 60,967 (46.1\%) all-time creators and 71,319 (53.9\%) active users. Each all-time creator created 1.12 projects on average. These creators were only interested in creating their own projects and sought funds. Interestingly, the average number of created projects per all-time creator reveals that these creators created just one or two projects. However, each of 71,319 active users created 1.25 projects and backed 6.45 projects on average. These active users created a little more projects than all-time creators, and backed many other projects.

A follow-up question is ``Do experienced users achieve a higher project success rate?''. We measured experience of a user based on when they create a project after joining Kickstarter. Figure~\ref{fig:interval} shows cumulative distribution functions (CDFs) of intervals between user joined date and project creation date in successful projects and failed projects. As we expected, successful projects had longer intervals. We conjecture that since users with longer intervals become more experienced and familiar with Kickstarter platform, their projects have become successful with a higher probability.

\subsection{Analysis of Projects}
\label{sec:analysis}

\begin{table}%[h]
\tbl{Statistics of Kickstarter projects.\label{table:quick-view-dataset}}
{
\centering
\small
\begin{tabular}{lrrr}
 & \textbf{Success} & \textbf{Failure} & \textbf{Total} \\ \hline
\begin{tabular}[c]{@{}l@{}}Percentage (\%)\end{tabular} & 46 & 54 & 100 \\
\begin{tabular}[c]{@{}l@{}}Classified project count\end{tabular} & 69,448 & 82,160 & 151,608 \\
\begin{tabular}[c]{@{}l@{}}Duration (days)\end{tabular} & 33.21 & 36.2 & 34.83 \\
%Minimum & 1 & 1 & 1 \\
Project Goal (USD) & 8,364.34 & 35,201.89 & 22,891.15 \\
Final money pledged (USD) & 16,027.96 & 1,454.18 & 8,139.37 \\
Number of images & 4.63 & 3.37 & 3.95 \\
Number of videos & 1.18 & 0.93 & 1.04 \\
Number of FAQs & 0.84 & 0.39 & 0.6 \\
Number of rewards & 9.69 & 7.49 & 8.5 \\
Number of updates & 9.59 & 1.59 & 5.26 \\
Number of project comments & 77.52 & 2.45 & 36.89 \\
Facebook connected (\%) & 61.00 & 59.00 & 60.00 \\
Number of FB friends & 583.48 & 395.15 & 481.54 \\
Number of backers & 211.16 & 19.34 & 107.33 \\ \hline
\end{tabular}
}
%\vspace{-5pt}
\end{table}

So far we analyzed collected user profiles. Now we turn to analyze Kickstarter projects. Interesting research questions are: How many projects have been created over time? What percent of projects has been successfully funded? Can we observe clearly different properties between successfully funded projects and failed projects? To answer these questions, we analyzed Kickstarter project dataset presented in Table~\ref{table:dataset}.

\begin{figure}%[h]
\centerline{
	\includegraphics[width=0.7\textwidth]{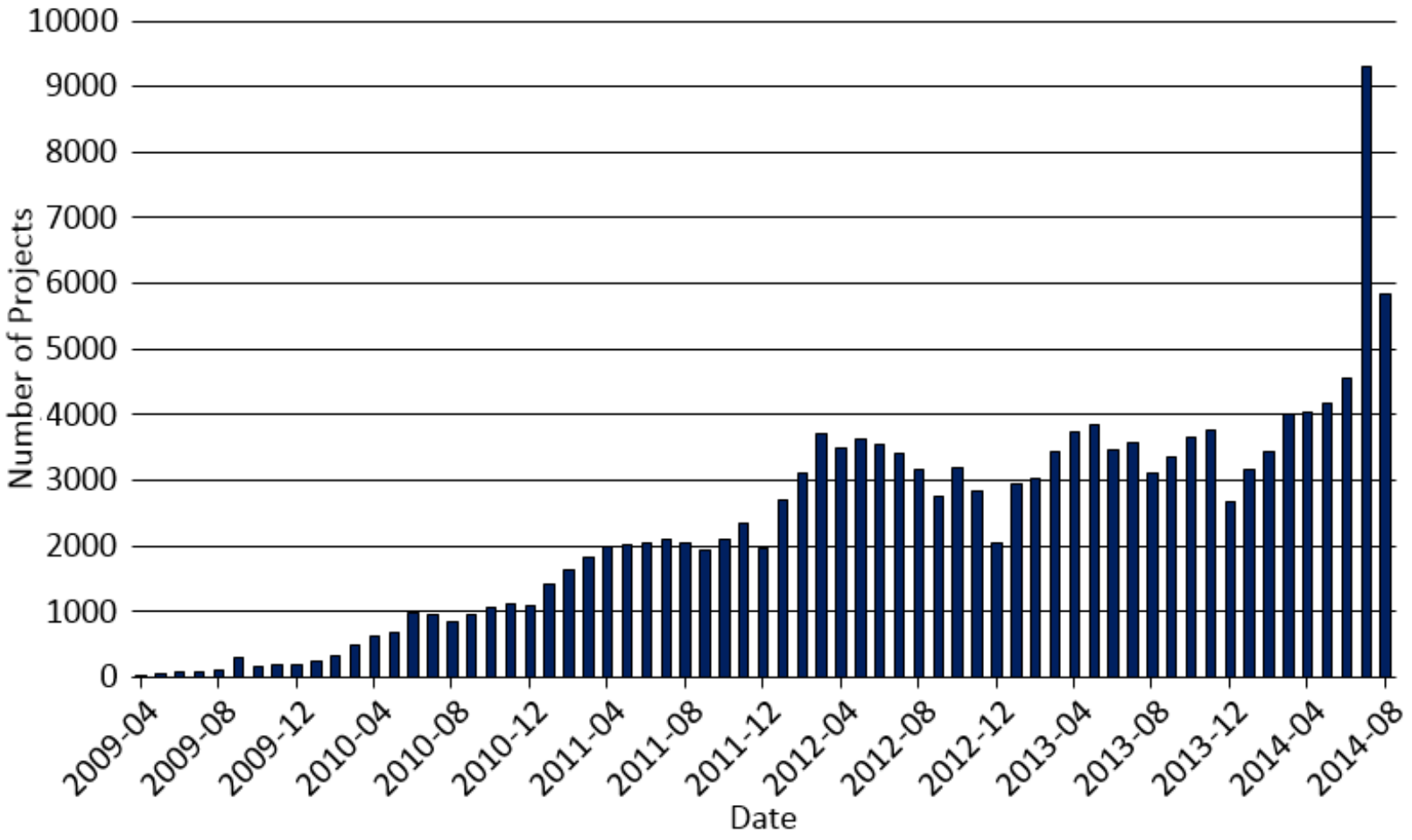}
}
\caption{Number of created projects per month has been increased over time with some fluctuation.}
\label{fig:number-projects-over-time}
%\vspace{-10pt}
\end{figure}

\smallskip
\noindent\textbf{Number of projects and project success rate over time.} Figure~\ref{fig:number-projects-over-time} shows how the number of projects has been changed over time. Overall, the number of created projects per month has been increased over time with some fluctuation. Interestingly, lower number of projects in December of each year (e.g., 2011, 2012 and 2013) has been created. Another interesting observation was that the largest number of projects (9,316 projects) were created in July 2014. The phenomena would be related to the number of newly joined users per month shown in Figure~\ref{fig:change-date} in which less number of users joined Kickstarter during Winter season, especially in December in each year, and many users joined in July 2014.

\begin{figure}%[!ht]
    \centering
    \begin{floatrow}
      \ffigbox[\FBwidth]
      {
      		\caption{Project success rate in each month.}
      		\label{fig:change-s-rate}
      }
      {
	  		\includegraphics[width=0.43\textwidth, height=1.6in]{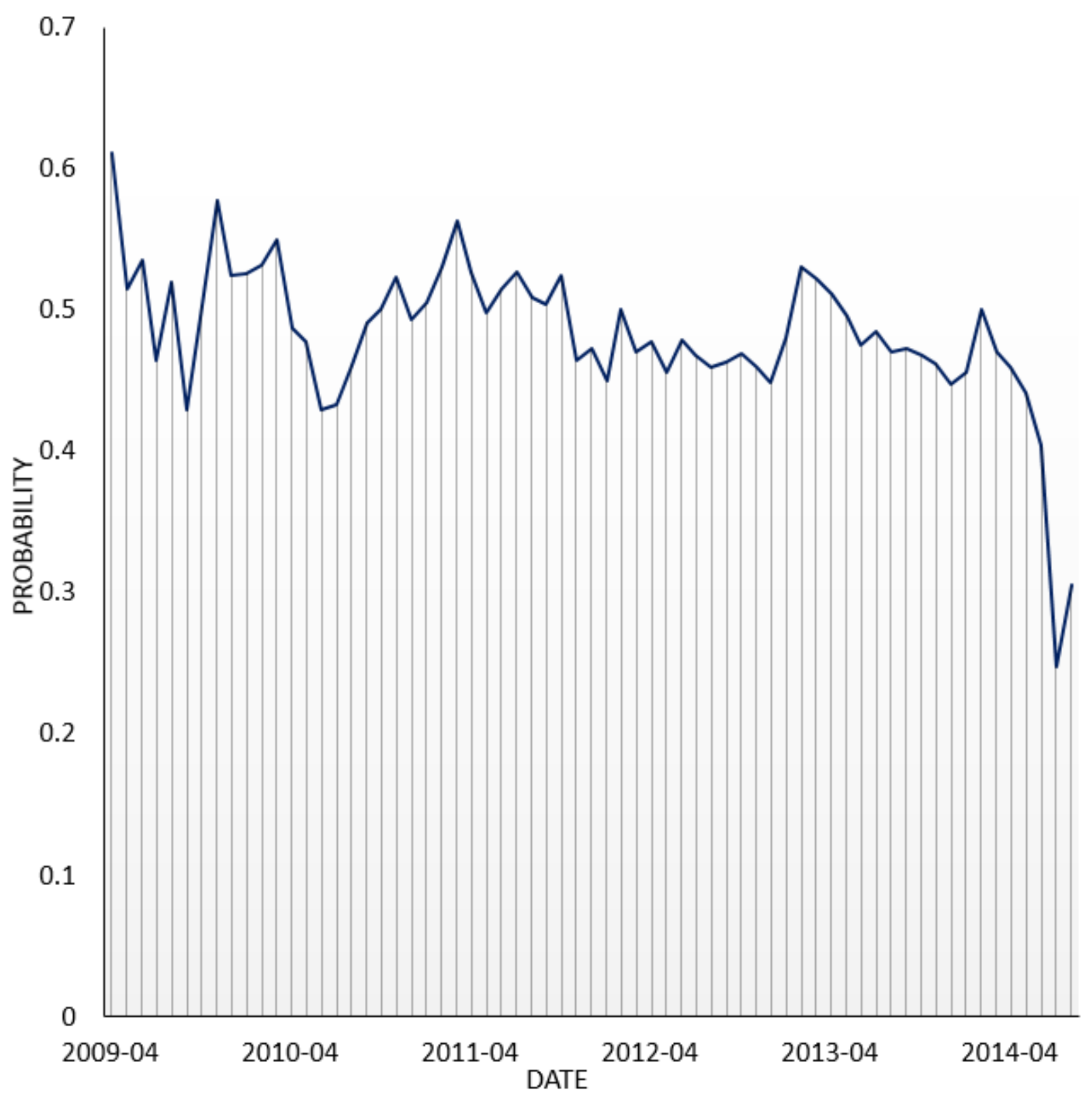}
      }
      \ffigbox[\FBwidth]
      	{	
      		\caption{Project success and failure rates according to a duration that more than 1,000 projects has.}
      		\label{fig:duration-rate}
      	}
      	{
        		\includegraphics[width=0.43\textwidth]{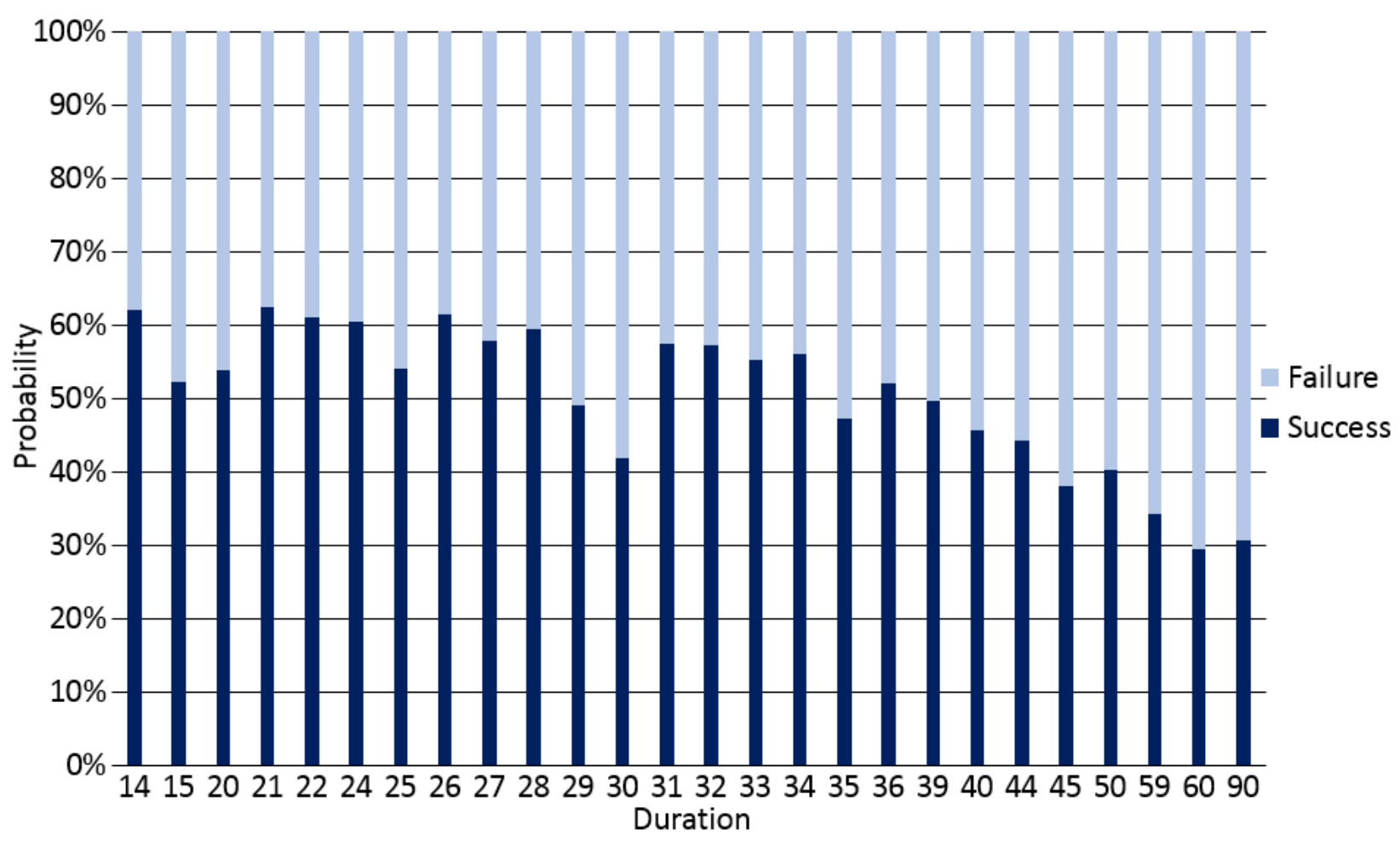}
      	}	
    \end{floatrow}
\end{figure}

Next, we are interested in analyzing how project success rate has been changed over time. We grouped projects by their launched year and month. Interestingly, the success rate has been fluctuated and overall project success rate in each month has been decreased over time as shown in Figure~\ref{fig:change-s-rate}. In July 2014, the success rate was dramatically decreased. We conjecture that since many users joined Kickstarter in July 2014, these first-time project creators caused the sharp decrease of success rate.

\smallskip
\noindent\textbf{Statistics of successful projects and failed projects.} Next, we analyze statistics of successful projects and failed projects. Table~\ref{table:quick-view-dataset} presents the statistics of Kickstarter projects. Overall, percentage of the successful projects in our dataset is about 46\%. In other words, 54\% of all projects was failed. We can clearly observe that the successful projects had shorter project duration, lower funding goal, more active engagements and larger number of social network friends than failed projects.

\begin{figure}[t]
    \centering
    \begin{floatrow}
      \ffigbox[\FBwidth]
      {
      		\caption{Number of Projects according to a duration that more than 1,000 projects has.}
      		\label{fig:duration-num-projects}
      }
      {
	  		\includegraphics[width=0.465\textwidth, height=1.8in]{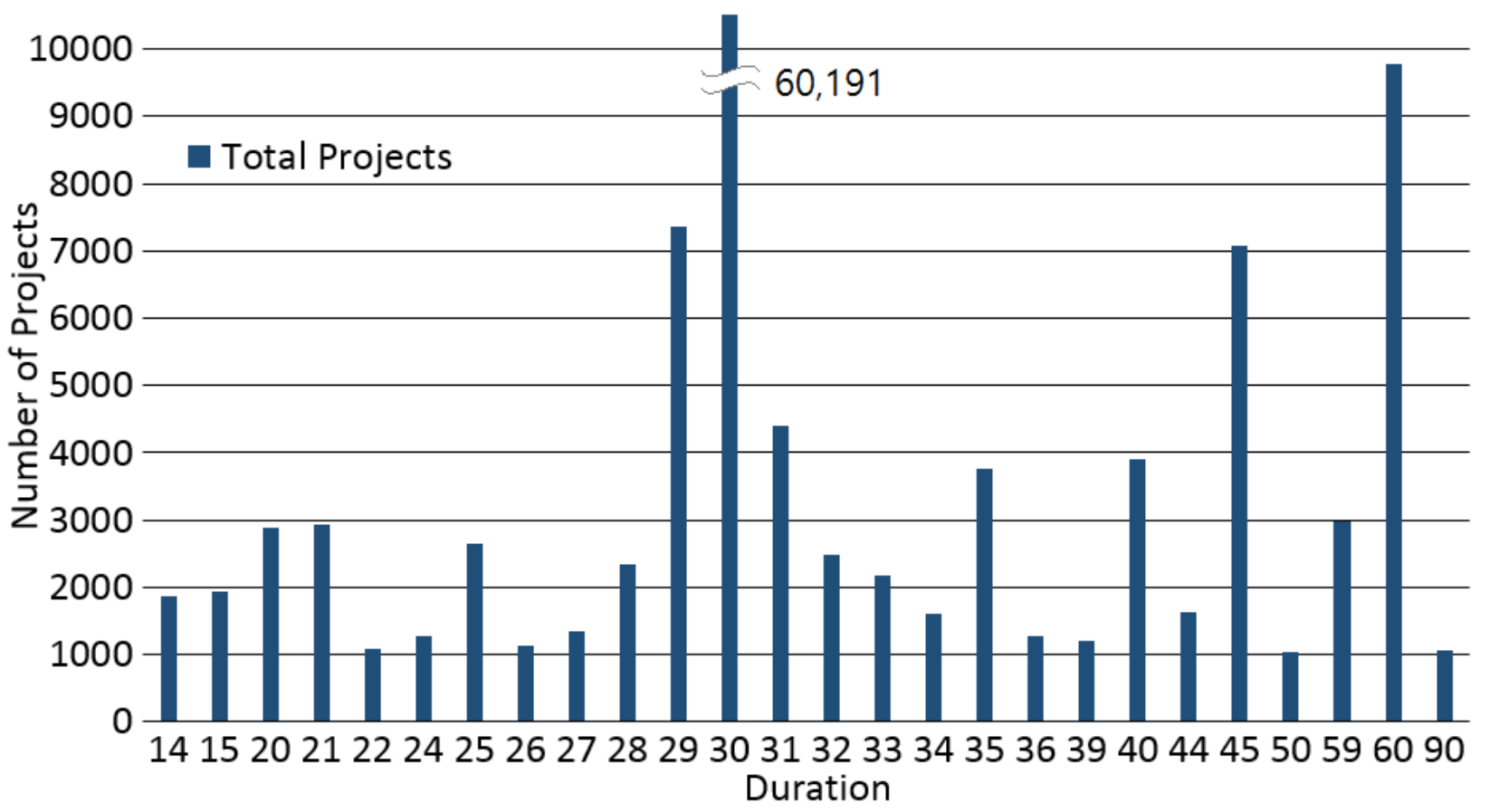}
      }
      \ffigbox[\FBwidth]
      	{	
      		\caption{Project success rate under each of 15 categories.}
      		\label{fig:category-s-rate}
      	}
      	{
        		\includegraphics[width=0.465\textwidth]{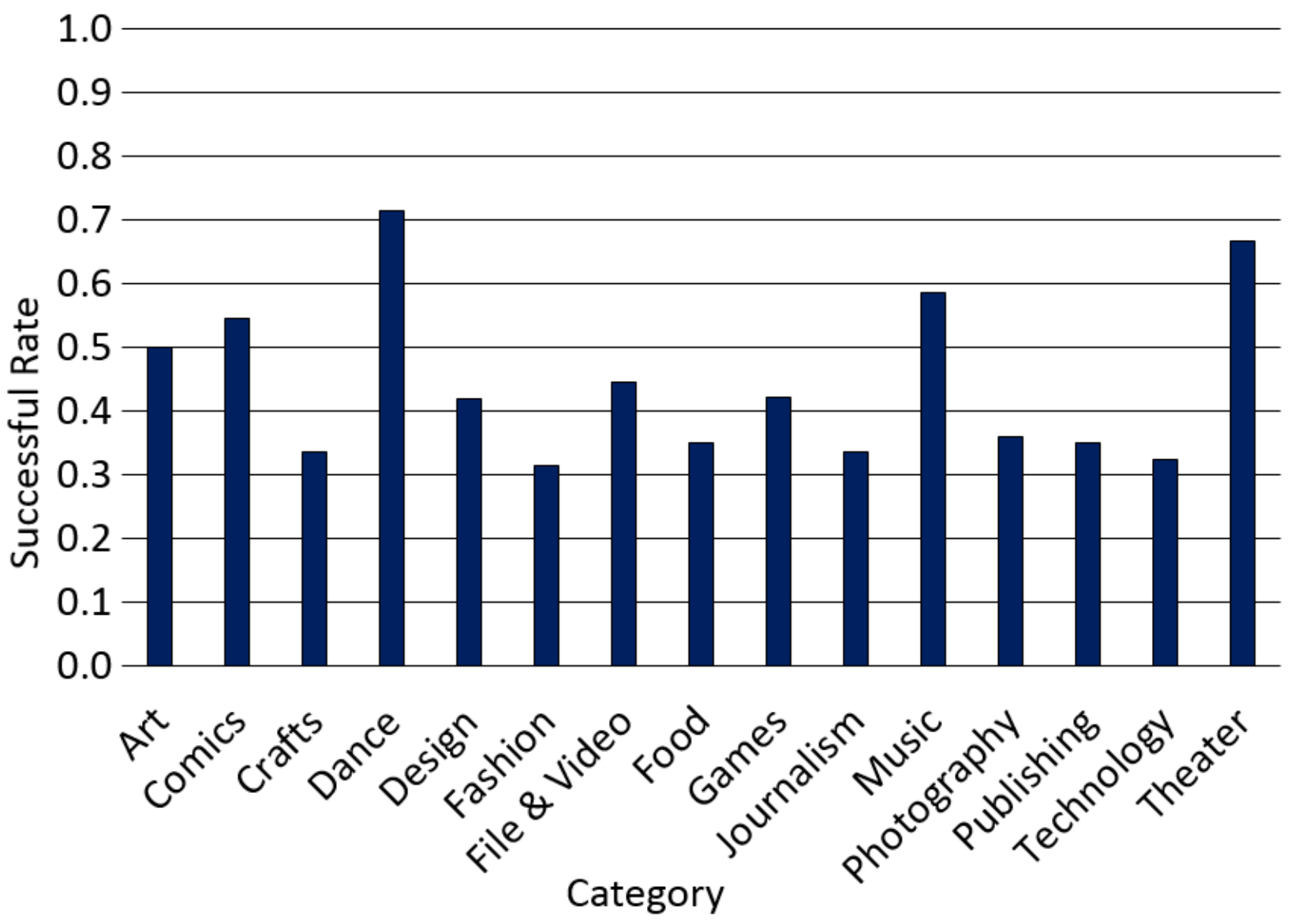}
      	}	
    \end{floatrow}
\end{figure}

Figure~\ref{fig:duration-rate} shows more detailed information about how project success rate was changed when a project duration was increased. This figure clearly shows that project success rate was higher when a projet duration was shorter. Intuitively, people may think that longer project duration would be helpful to get more fund, but this analysis reveals the opposite result. To show how many projects have what duration, we plotted Figure~\ref{fig:duration-num-projects}. 39.7\% (60,191 projects) of all projects had 30 day duration and then 6.5\% (9,784 projects) of all projects had 60 day duration. We conjecture that since 30 day duration is the default duration on Kickstarter, many users just chose 30 day duration for their projects. %Note that since some of Kickstarter projects have been created in no-US countries like England with non-US currency like Pound, we converted non-US currency to US dollar based on the exchange rate as of November 2014.

While the average project goal of successful projects was 3 times less than failed projects, the average pledged money of successful projects was 10 times more than failed projects. Project creators of successful projects spent more time to make better project description by adding a larger number of images, videos, FAQ and reward types. The creators also frequently updated their projects. Interestingly, project creators of the successful projects had a larger number of Facebook friends. It means that the creators' Facebook friends might help for their project success by backing the projects or spreading information of the projects to other people \cite{mollick:2014}.

When a user creates a project on Kickstarter, she can choose a category of the project. Does a category of a project affect a project success rate? To answer this question, we analyzed project success rate according to each category. As you can see in Figure~\ref{fig:category-s-rate}, projects in Dance, Music, Theater, Comics and Art categories achieved between 50\% and 72\% success rate which is greater than the average success rate of all projects (again, 46\% success rate).

\begin{figure}
    \centering
    \begin{floatrow}
      \ffigbox[\FBwidth]
      {
              \caption{Distribution of projects in the world.}
              \label{fig:loc-world}
      }
      {
              \includegraphics[width=0.33\textwidth]{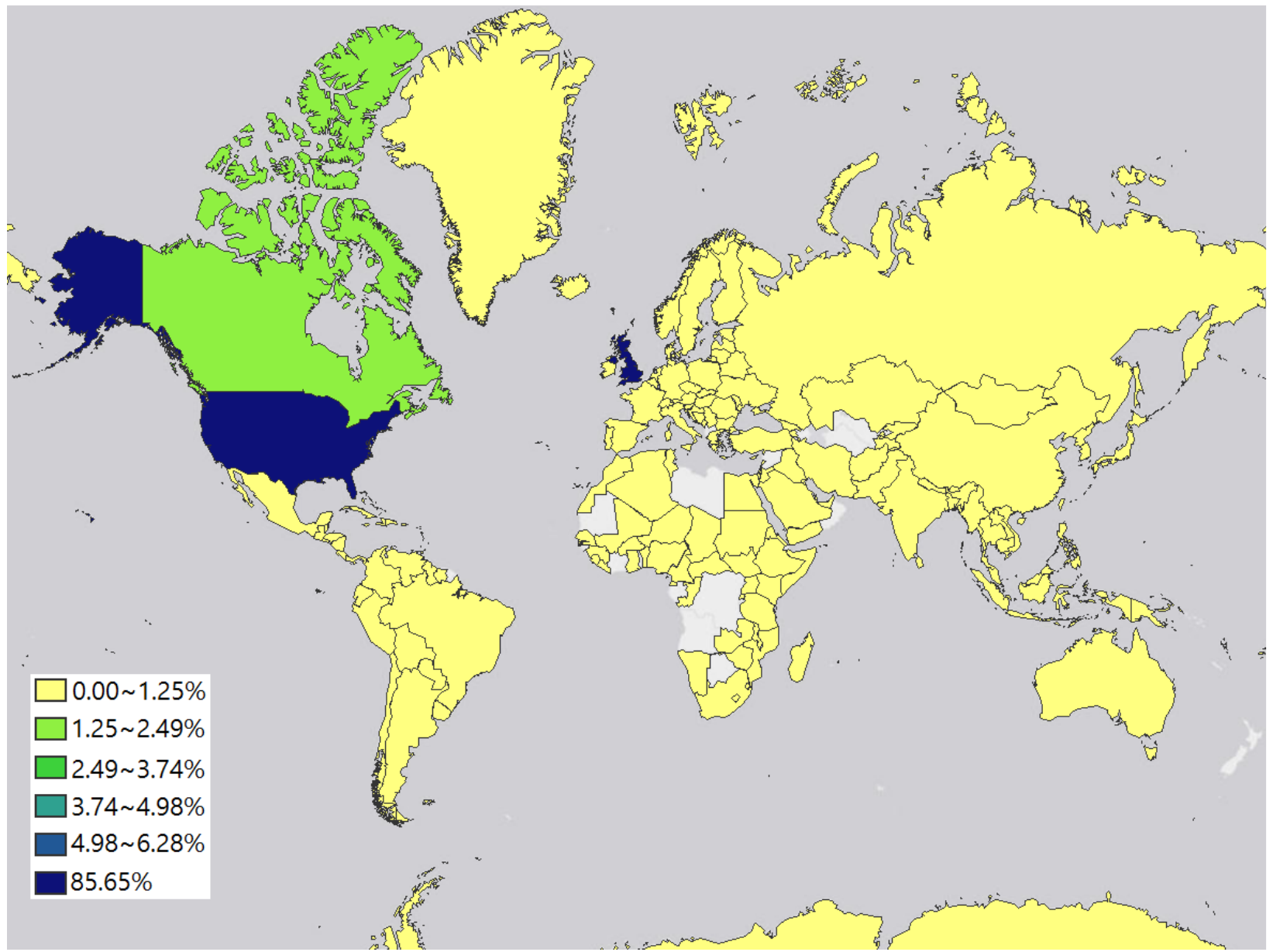}
      }
      \ffigbox[\FBwidth]
          {
              \caption{Distribution of projects in US.}
              \label{fig:loc-us-num}
          }
          {
                \includegraphics[width=0.6\textwidth]{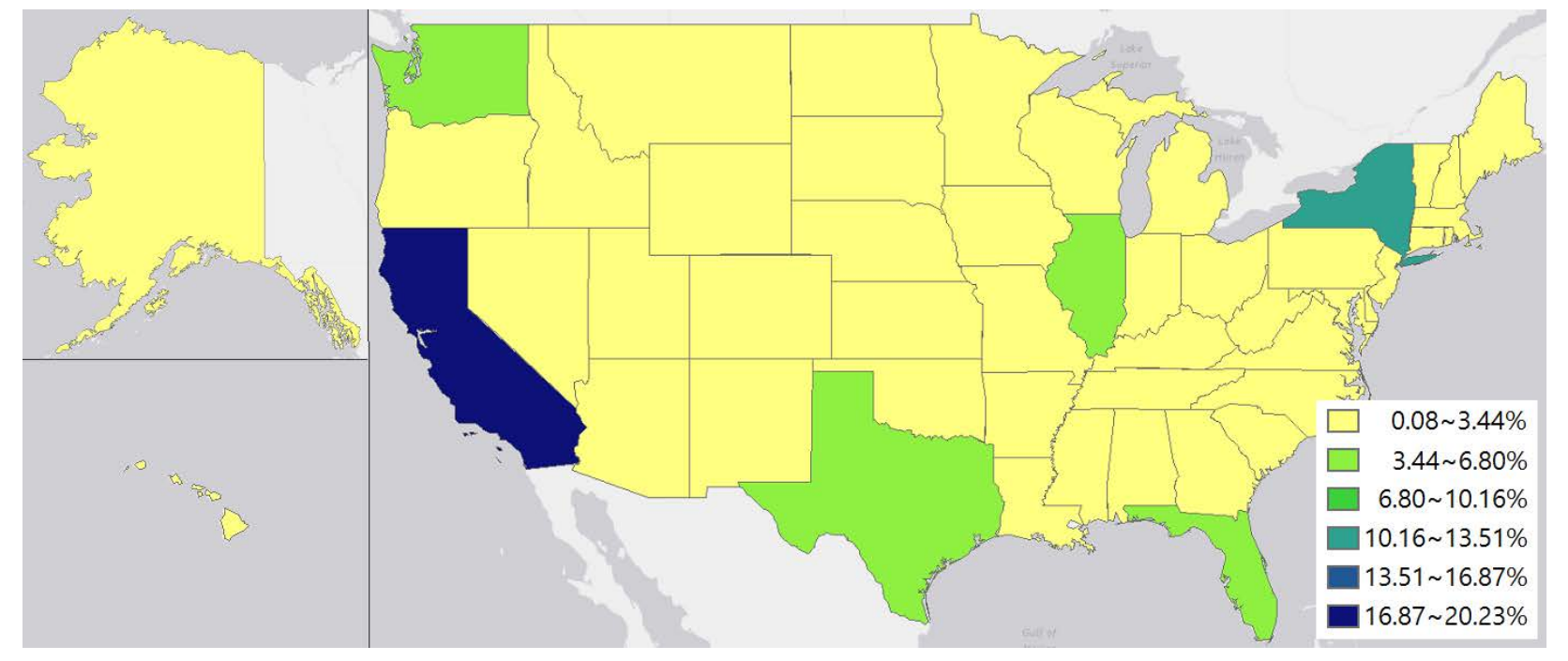}
          }
    \end{floatrow}
\end{figure}

\smallskip
\noindent\textbf{Location.} A user can add location information when she creates a project. We checked our dataset to see how many projects contain location information. Surprisingly, 99\% project pages contained location information. After extracting the location information from the projects, we plotted distribution of projects on the world map in Figure~\ref{fig:loc-world}. 85.65\% projects were created in US. The next largest number of projects were created in the United Kingdom (6.23\%), Canada (2.20\%), Australia (1\%)and Germany (0.92\%). Overall, the majority of projects were created in the western countries. The project distribution across countries makes sense because initially only US based projects on Kickstarter were created, and then the company allowed users in other countries to launch projects since October 2012. Since over 85\% projects were created in US, we plotted distribution of the projects on US map in Figure~\ref{fig:loc-us-num}. Top 5 states are California (20.23\%), New York (12.93\%), Texas (5.45\%), Florida (4.57\%) and Illinois (4.03\%). This distribution mostly follows population of each state.

\begin{figure}%[h]
\centerline {
	\includegraphics[height=1.5in]{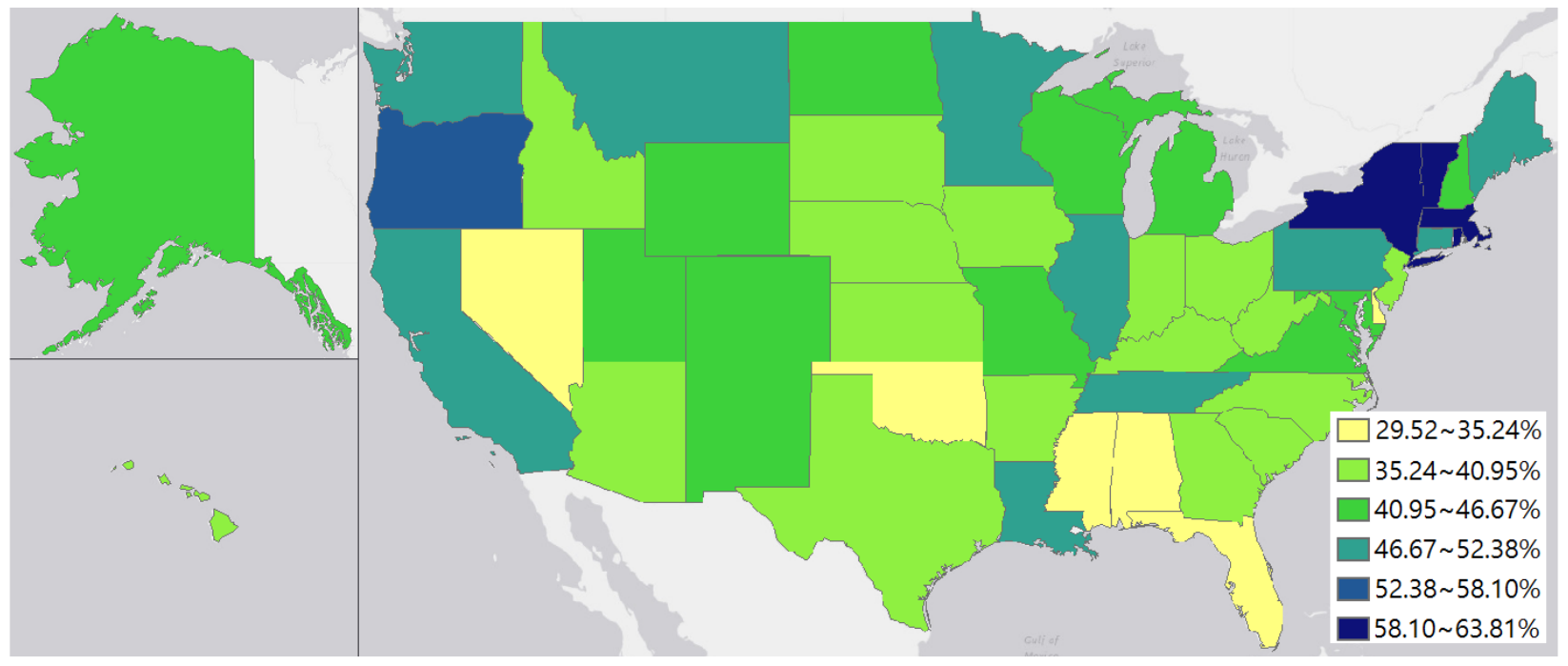}
}
\caption{Project success rate across states in US.}
\label{fig:loc-us-success}
%\vspace{-10pt}
\end{figure}

A follow-up question is how project distribution across states in US is related to projects success rate. To answer this question, we plotted project success rate of each state in Figure~\ref{fig:loc-us-success}. Top 5 states with the highest success rate are Vermont (63.81\%), Massachusetts (58.49\%), New York (58.46\%), Rhode Island (58.33\%) and Oregon (53.56\%). Except New York state, small number of projects were created in the four states. To make a concrete conclusion, we measured Pearson correlation between distribution of projects and project success rate. The correlation value was 0.25 which indicates that they are not significantly correlated.

\smallskip
\noindent\textbf{Analysis of Kickstarter Temporal Data.} As we presented in Table~\ref{table:dataset}, we collected temporal data of 74,053 projects (e.g., daily pledged money and daily increased number of backers). Using these temporal data, we analyzed what percent of total pledged money and what percent of backers each project got over time after launching a project. Since each project has different duration (e.g., 30 days or 60 days), first, we converted each project duration to 100 states (time slots). Then, in each state, we measured percent of pledged money and number of backers.

\begin{figure}
\centerline {
	\includegraphics[height=2in]{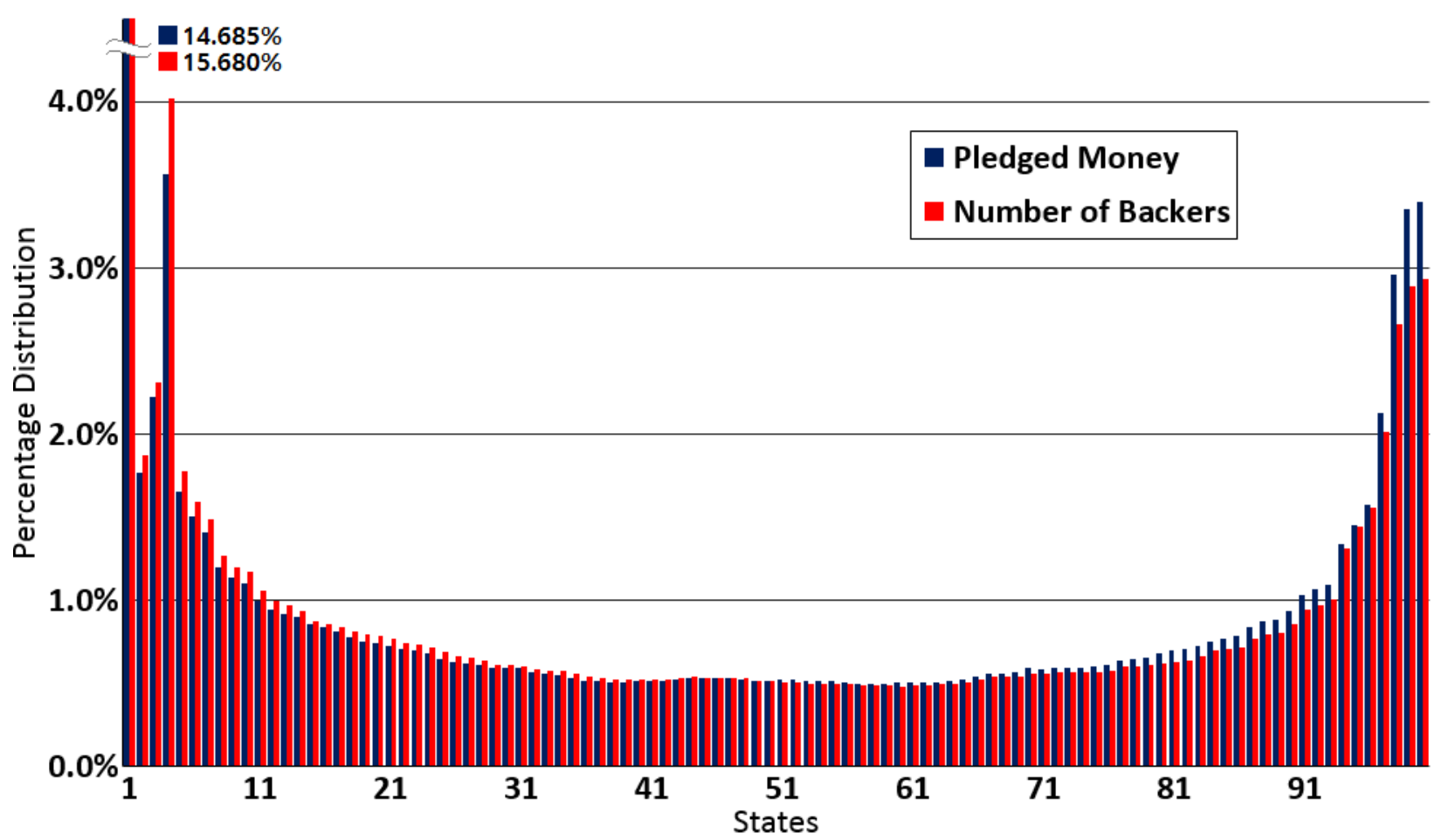}
}
\caption{Percentage distribution of pledged money and number of backers per state.}
\label{fig:avg-pledgedmoney-perState}
%\vspace{-10pt}
\end{figure}

Figure~\ref{fig:avg-pledgedmoney-perState} shows the percentage distribution of pledged money and number of backers per state over time. One of the most interesting observations is that the largest amount of money was pledged in the beginning and end of a project. For example, 14.69\% money was pledged and 15.68\% backers were obtained in the first state. Other researchers also observed the same phenomena in smaller datasets \cite{Kuppuswamy,lu:2014}.

Another interesting observation is that there is another spike after the first spike in the beginning of project durations. We conjecture that the first spike was caused by a project creator's family and friends who backed the project \cite{thundering}, and the second spike was caused by other users who noticed the project and heard of a trend of the project.

The other interesting observation is that after 60th state, the number of backers and the number of pledged money have been exponentially increased. Especially, people rushed investing a project, as a project was heading to the end of the project duration. The phenomenon is called the Deadline effect \cite{roth:1988},\cite{yildiz:2004}. Even amount of invested money has been increased more quickly than the number of backers. This may indicate that people tend to purchase more expensive reward item. They may want to make sure a project become successful, achieving higher amount of pledged money than a project goal\footnote{Kickstarter has an All-or-Nothing policy. If a project reaches at or over its goal, its creator will receive pledged fund. Otherwise, the project creator will receive nothing.}. In another case, they knew that other people already supported the project with a large amount of money which motivated them to back the project with high trust.

\section{Features and Experimental Settings}
\label{sec:features}
In the previous section, we analyzed behaviors of Kickstarter users and characteristics of projects. Based on the analysis, in this section we propose features which will be used to develop a project success predictor and an expected funding range predictor. We also describe our experimental settings which are used in Sections~\ref{sec:success} and~\ref{sec:range}.

\subsection{Features}
We extracted 49 features from our collected datasets presented in Table~\ref{table:dataset}. Then, we grouped the features to 4 types: (i) project features; (ii) user features; (iii) temporal features; and (iv) Twitter features.

\subsubsection{Project Features}
From a project page, we generated 11 features as follows:

\begin{itemize}

\item Project category, duration, project goal, number of images, number of videos, number of FAQs, and number of rewards.
\item SMOG grade of reward description: To estimate the readability of the all rewards text.
\item SMOG grade of main page description: To estimate the readability of the main page description of a project.
\item Number of sentences in reward description.
\item Number of sentences in the main description of a project.
\end{itemize}

The SMOG grade estimates the years of education needed to understand a piece of writing \cite{McLaughlin1969}. The higher SMOG grade indicates that project and reward descriptions were written well. To measure SMOG grade, we used the following formula:
%\begin{equation} \label{eq:smog}
\[
1.043\sqrt{|polysyllables| \times \frac{30}{|sentences|}}  + 3.1291
\]
%\end{equation}
, where the number of Polysyllables is the count of the words of 3 or more syllables.

\subsubsection{User Features}
From a user profile page and the user's previous experience, we generated 28 features as follows:

\begin{itemize}
\item Distribution of the backed projects under the 15 main categories (15 features): what percent of projects belongs to each main category.
\item Number of backed projects, number of created projects in the past, number of comments that a user made in the past, number of websites linked in a user profile, and number of Facebook friends that a user has.
\item Is each of Facebook, YouTube and Twitter user pages connected? (3 features)
\item SMOG grade of bio description, and Number of sentences in a bio description.
\item Interval (days) between a user's Kickstarter joined date and a project's launched date.
\item Success rate of the backed projects by a user.
\item Success rate of the projects created by a user in the past.
\end{itemize}

\subsubsection{Temporal Features}
As we mentioned in Section~\ref{sec:dataset}, we collected 74,053 projects' temporal data consisting of daily pledged money and number of daily increased backers. First, we converted these temporal data points (i.e., daily value) to cumulated data points. For example, if a project's daily pledged money for 5 days project duration are 100, 200, 200, 100 and 200, cumulated data point in each day will be 100, 300, 500, 600 and 800. Since each project has various duration, we converted a duration to 100 states (time slots). Then, we normalized cumulated data points by 100 states. Finally, we generated two time-series features:
\begin{itemize}
\item Cumulated pledged money over time.
\item Cumulated number of backers over time.
\end{itemize}

\subsubsection{Twitter Features}
As we mentioned in Section~\ref{sec:dataset}, 17,908 users linked their Twitter home pages to their Kickstarter user pages. From our collected Twitter dataset, we generated 8 features as follows:

\begin{itemize}
\item Number of tweets, Number of followings, Number of followers and Number of favorites.
\item Number of lists that a user has been joined in.
\item Number of tweets posted during active project days (e.g., between Jan 1, 2014 and Jan 30, 2014).
\item Number of tweets containing word ``Kickstarter'' posted during active project days.
\item SMOG grade of aggregated tweets which are posted during active project days.
\end{itemize}

The first five features were used for any project created by a user. The rest three features were generated for each project since each project was active in different time period.

Finally, we generated 49 features from a project and a user who created the project.

\subsection{Experimental Settings}
We describe our experimental settings which are used in the following sections for predicting project success and expected pledged money range.

\smallskip
\noindent\textbf{Datasets.} In the following sections, we used three datasets presented in Table~\ref{table:number-instance}. Each dataset consists of a different number of projects and corresponding user profiles as we described in Section~\ref{sec:dataset}. Two datasets (KS Static + Twitter, and KS Static + Temporal + Twitter) contained Twitter user profiles as well.

We extracted 39 features from KS Static dataset (i.e., project features and user features), 47 features from KS Static + Twitter dataset (i.e., project features, user features and Twitter features), and 49 features from KS Static + Temporal + Twitter (i.e., all four feature groups). Note that in this subsection we presented the total number of our proposed features before applying feature selection.

\begin{table}%[h]
	\tbl{Three datasets which were used in experiments.\label{table:number-instance}}
	{
		\small
		\centering
		\begin{tabular}{|l|r|r|}
			\hline
			\textbf{Datasets}            & \textbf{$|$Projects$|$} & \textbf{$|$Features$|$} \\ \hline
			KS Static & 151,608              & 39                  \\ \hline
			KS Static + Twitter  & 21,028               & 47                  \\ \hline
			%KS Static + Temporal      & 74,053               & 41                  \\ \hline
			KS Static + Temporal + Twitter & 11,675               & 49                  \\ \hline
		\end{tabular}
	}
\end{table}

\smallskip
\noindent\textbf{Predictive Models.} Since each classification algorithm might perform differently in our dataset, we selected 3 well-known classification algorithms: Naive Bayes, Random Forest, AdaboostM1 (with Random Forest as the base learner). We used Weka implementation of these algorithms \cite{Hall:2009}.

\smallskip
\noindent\textbf{Feature Selection.} To check whether the proposed features were positively contributing to build a good predictor, we measured $\chi^{2}$ value \cite{657137} for each of the features. The larger the $\chi^{2}$ value is, the higher discriminative power the corresponding feature has. The feature selection results are described in following sections.

\smallskip
\noindent\textbf{Evaluation.} We used Accuracy as the primary evaluation metrics and Area under the ROC Curve (AUC) as the secondary metrics, and then built and evaluated each predictive model (classifier) by using 5-fold cross-validation.

\section{Predicting Project Success}
\label{sec:success}
Based on the features and experimental settings, we now develop and evaluate project success predictors.

\subsection{Feature Selection}
First of all, we conducted $\chi^{2}$ feature selection to check whether the proposed features were all significant features. Since we had three datasets, we applied feature selection for each dataset. All features in KS Static dataset had positive distinguishing power to determine whether a project will be successful or not. But, in both of KS Static + Twitter dataset and KS Static + Temporal + Twitter, ``Is each of Facebook, YouTube and Twitter user pages connected'' features were not positively contributing, so we excluded them. Overall, some of project features (e.g., category, goal and number of rewards), some of user features (e.g., number of backed projects, success rate of backed projects, number of comments), some of Twitter features (e.g. number of lists, number of followers and number of favorites), and all temporal features were the most significant features.

\subsection{Experiments}
Our experimental goal is to develop and evaluate project success predictors. We build project success predictors by using each of the three datasets and evaluate performance of the predictors.

\smallskip
\noindent\textbf{Using KS Static dataset.} The first task was to test whether only using Kickstarter static features (i.e., project and user features) would achieve good prediction results. To conduct this task, we converted Kickstarter static dataset consisting of 151,608 project profiles and user profiles to feature values. Then, We developed project success predictors based on each of 3 classification algorithms -- Naive Bayes, Random Forest and AdaboostM1. Finally, we evaluated each predictor by using 5-fold cross-validation. Table~\ref{table:predict-success} shows experimental results of three project success predictors based on Kickstarter static features. AdaboostM1 outperformed the other predictors, achieving 76.4\% accuracy and 0.838 AUC. This result was better than 54\% accuracy of a baseline which was measured by a percent of the majority class instances in Kickstarter static dataset (54\% projects were unsuccessful). This result was also better than the previous work in which 68\% accuracy was achieved \cite{Greenberg:2013}.

\begin{table}%[h]
	\tbl{Experimental results of three project success predictors based on Kickstarter static features.\label{table:predict-success}}
	{
		\centering
		\small
		\begin{tabular}{|c|cc|}
			\hline
			Classifier & Accuracy & AUC   \\ \hline
			%J48                 & 73.7\%   & 0.731     \\
			Naive Bayes         & 67.3\%   & 0.750     \\
			Random Forest       & 75.2\%   & 0.827     \\
			AdaboostM1          & \textbf{76.4\%}  & \textbf{0.838} \\ \hline
			%SMO                 & 70.9\%   & 0.704        \\ \hline
		\end{tabular}
	}
	%\vspace{-10pt}
\end{table}

\begin{table}%[h]
\tbl{Project success predictors based on Kickstarter static features vs. based on Kickstarter static features and Twitter features.\label{table:predict-success-twt}}
{
\small
\centering
\begin{tabular}{|c|cc|}
\hline
Classifier & Accuracy & AUC \\ \hline
\multicolumn{3}{|c|}{\textbf{Kickstarter}}                                       \\ \hline
Naive Bayes & 60.3\% & 0.722         \\
Random Forest          & 72.8\% & 0.790         \\
AdaboostM1  & \textbf{73.9\%} & \textbf{0.798} \\ \hline
\multicolumn{3}{|c|}{\textbf{Kickstarter + Twitter}}                             \\ \hline
Naive Bayes & 56.5\% & 0.724 \\
Random Forest          & 73.4\% & 0.800 \\
AdaboostM1  & \textbf{75.7\%} & \textbf{0.826} \\ \hline
\end{tabular}
}
%\vspace{-5pt}
\end{table}

\smallskip
\noindent\textbf{Using KS Static + Twitter dataset.} What if we add Twitter features to Kickstarter static features? Can we even improve performance of project success predictors? To answer these questions, we compared performance of predictors without Twitter features with performance of predictors with Twitter features. In this experiment, we extracted Kickstarter static features from 21,028 projects and corresponding user profiles, and Twitter features from corresponding Twitter user profiles. As you can see in Table~\ref{table:predict-success-twt}, AdaboostM1 classifier with Twitter features achieved 75.7\% accuracy and 0.826 AUC, increasing accuracy and AUC of AdaboostM1 classifier without Twitter features by 2.5\% (= $\frac{75.7}{73.9} - 1$) and 3.5\% (= $\frac{0.826}{0.798} - 1$), respectively.

\begin{figure}
	\centerline{
		\includegraphics[width=0.7\textwidth]{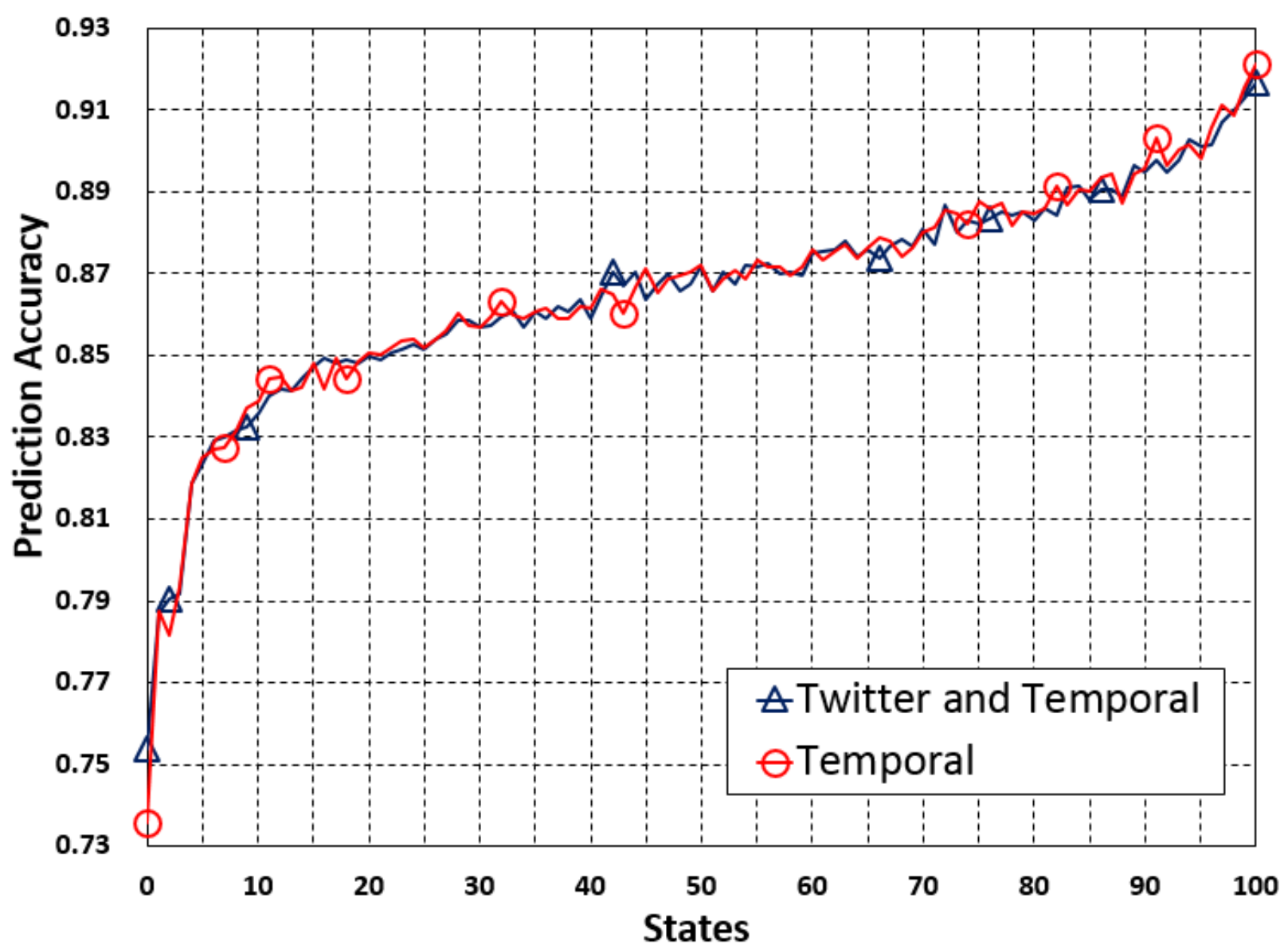}
	}
	\caption{Project success prediction rate of predictors based on Kickstarter static and temporal features with/without Twitter features.}
	\label{fig:TEM-TWT-SorF}
	%\vspace{-15pt}
\end{figure}

\smallskip
\noindent\textbf{Using KS Static + Temporal + Twitter dataset.} What if we replace Twitter features with Kickstarter temporal features? Or what if we use all features including Kickstarter static, temporal and Twitter features? Would using all features give us the best result?  To answer these questions, we used KS Static + Temporal + Twitter dataset consisting of 11,675 project profiles, corresponding user profiles, Twitter profiles and project temporal data. Since each project has a different project duration, we converted each project duration to 100 states (time slots). Then we calculated temporal feature values in each state. Finally, we developed 100 predictors based on KS Static + Temporal features and 100 predictors based on KS Static + Temporal + Twitter features (each predictor was developed in each state). Note that in the previous experiments AdaboostM1 consistently outperformed the other classification algorithms, so used AdaboostM1 for this experiment. Figure~\ref{fig:TEM-TWT-SorF} shows two project success predictors' accuracy in each state. In the beginning, KS Static + Temporal + Twitter features based predictors were slightly better than KS Static + Temporal features based predictors, but both of approaches performed similarly after 3rd state because temporal features became more significant. Overall, accuracy of predictors has been sharply increased until 11th state and then consistently increased until the end of a project duration. In 10th state (i.e., in the first 10\% duration), the predictors achieved 83.6\% accuracy which was increased by 11\% (= $\frac{83.6}{75.3} - 1$) compared with 75.3\% accuracy when a state was 0 (i.e., without temporal features). The more a state value increased, the higher accuracy a predictor achieved.

In summary, we developed project success predictors with various feature combinations. A project success predictor based on Kickstarter static features achieved 76.4\% accuracy. Adding social media features increased the prediction accuracy by 2.5\%. Adding temporal features consistently increased the accuracy. The experimental results confirmed that it is possible to predict a project's success when a user creates a project, and we can increase a prediction accuracy further with early observation after launching the project.

\section{Predicting an Expected Pledged Money Range of a Project}
\label{sec:range}
So far we have studied predicting whether a project will be successful or not. But a project's success depends on a project goal and pledged money. If pledged money is equal to or greater than a project goal, the project will be successful. On the other hand, even though a project received a lot of pledged money (e.g., \$99,999) , if a project goal (e.g., \$100,000) is slightly larger than the pledged money, the project will be failed. Remember the All-or-Nothing policy. If we predict how much a project will get invested in advance, we can set up a realistic project goal and make the project successful. A fundamental research problem is ''Can we predict expected pledged money? or Can we predict a range of expected pledged money of a project?''. To our knowledge, no one has studied this research problem yet. In this section, we propose an approach to predict a range of expected pledged money of a project.

\subsection{Approach and Feature Selection}
In this section, our research goal is to develop predictive models which can predict a range of pledged money of a project. We defined the number of classes (categories) in two scenarios: (i) 2 classes; and (ii) 3 classes. In a scenario of 2 classes, we used a threshold, \$5,000. The first class is $\leq \$5,000$, and the second class is $> \$5,000$. In other words, if pledged money of a project is less than or equal to \$5,000, the project will belong to the first class. Likewise, in a scenario of 3 classes, we used two thresholds, \$100 and \$10,000. The first class is $\leq \$100$, the second class is $\$100<project\leq\$10,000$ and the third class is $> \$10,000$. Now we have the ground truth in each scenario.

Next, we applied feature selection to our datasets. In 2 classes, ``Is Youtube connected'' feature was not a significant feature in KS Static and KS Static + Temporal + Twitter datasets. ``Is Twitter connected'' feature was not a significant feature in KS Static + Twitter and KS Static + Temporal + Twitter datasets. In 3 classes, ``Is Twitter connected'' feature was not a significant feature in KS Static + Twitter and KS Static + Temporal + Twitter datasets.

\subsection{Experiments}
%As we mentioned in the previous subsection,
We conducted experiments in two scenarios -- prediction in (i) 2 classes and (ii) 3 classes.

\begin{table}[h]
\tbl{Experimental results of pledged money range predictors based on Kickstarter static features under 2 classes.\label{table:predict-class2-range}}
{
\small
\centering
\begin{tabular}{|c|cc|}
\hline
Classifier & Accuracy & AUC \\ \hline
%J48                 & 0.851 & 0.813                \\
Naive Bayes         & 75.9\% & 0.780                 \\
Random Forest       & 85.6\% & \textbf{0.906}                \\
AdaboostM1          & \textbf{86.5\%} & 0.901       \\ \hline
%SMO                 $ 0.802 & 0.646         \\ \hline
\end{tabular}
}
%\vspace{-15pt}
\end{table}

\smallskip
\noindent\textbf{Using KS Static dataset.} The first experiment was to predict a project's pledged money range by using KS Static dataset (i.e., generating the static features -- project features and user features). A use case is that when a user creates a project, this predictor helps the user to set up an appropriate goal. We conducted 5 fold cross-validation in each of the two scenarios. Table~\ref{table:predict-class2-range} shows experimental results in 2 classes. AdaboostM1 outperformed Naive Bayes and Random Forest, achieving 86.5\% accuracy and 0.901 AUC. When we compared our predictor's performance with the baseline -- 74.8\% accuracy (percent of the majority class, assuming selecting the majority class as a prediction result) --, our approach increased 11.5\% (= $\frac{86.5}{74.8} - 1$).

\begin{table}[h]
\tbl{Experimental results of pledged money range predictors based on Kickstarter static features under 3 classes.\label{table:predict-class3-range}}
{
\small
\centering
\begin{tabular}{|c|cc|}
\hline
Classifier & Accuracy & AUC  \\ \hline
%J48                 & 0.715 & 0.718                \\
Naive Bayes    & 49.4\% & 0.713                \\
Random Forest                  & 73.3\% & \textbf{0.817}                \\
AdaboostM1          & \textbf{74.2\%} & 0.811                \\ \hline
%SMO                 & 0.699 & 0.679                \\ \hline
\end{tabular}
}
%\vspace{-5pt}
\end{table}

We also ran another experiment in 3 classes. Table~\ref{table:predict-class3-range} shows experimental results. Again, AdaboostM1 outperformed the other classification algorithms, achieving 74.2\% accuracy and 0.811 AUC. When we compared its performance with the baseline -- 63.1\% --, it increased 17.6\% (= $\frac{74.2}{63.1} - 1$). Regardless of the number of classes, our proposed approach consistently outperformed than the baseline. The experimental results showed that it is possible to predict an expected pledged money range in advance.

\begin{table}[h]
\tbl{Experimental results of pledged money range predictors based on Kickstarter static features and Twitter features under 2 classes.\label{table:predict-class2-range-twt}}
{
\small
\centering
\begin{tabular}{|c|cc|}
\hline
Classifier & Accuracy & AUC  \\ \hline
\multicolumn{3}{|c|}{Kickstarter}                                                                       \\ \hline
Naive Bayes     & 70.6\% & 0.759               \\
Random Forest  & 81.4\% & 0.889                \\
AdaboostM1     & \textbf{82.5\%} & \textbf{0.896}    \\ \hline
\multicolumn{3}{|c|}{Kickstarter + Twitter}                                                             \\ \hline
Naive Bayes         & 70.7\% & 0.763                \\
Random Forest                  & 83.1\% & 0.904                \\
AdaboostM1          & \textbf{84.2\%} & \textbf{0.910}                 \\ \hline
\end{tabular}
}
%\vspace{-5pt}
\end{table}

\smallskip
\noindent\textbf{Using KS Static + Twitter dataset.} What if we add Twitter features? Will these improve a prediction accuracy? To answer this research question, we used KS Static + Twitter dataset in each of 2 classes and 3 classes. Experimental results under 2 classes and 3 classes are shown in Tables~\ref{table:predict-class2-range-twt} and~\ref{table:predict-class3-range-twt}, respectively. In case of 2 classes, AdaboostM1 with Twitter features increased 2.1\% (= $\frac{84.2}{82.5} - 1$) compared with a predictor without Twitter features, achieving 84.2\% accuracy and 0.91 AUC. In case of 3 classes, AdaboostM1 with Twitter features also increased 1.8\% (= $\frac{77.2}{75.8} - 1$) compared with a predictor without Twitter features, achieving 77.2\% accuracy and 0.843 AUC. The experimental results confirmed that adding Twitter features improved prediction performance.

\begin{table}%[h]
\tbl{Experimental results of pledged money range predictors based on Kickstarter static features and Twitter features under 3 classes.\label{table:predict-class3-range-twt}}
{
\small
\centering
\begin{tabular}{|c|cc|}
\hline
Classifier & Accuracy & AUC  \\ \hline
\multicolumn{3}{|c|}{\textbf{Kickstarter}}                                                              \\ \hline
Naive Bayes         & 48.6\%  & 0.677               \\
Random Forest       & 74.2\%  & 0.829                \\
AdaboostM1          & \textbf{75.8\%}  & \textbf{0.830}                \\ \hline
\multicolumn{3}{|c|}{\textbf{Kickstarter + Twitter}}                                                    \\ \hline
Naive Bayes         & 48.8\%   & 0.668               \\
Random Forest       & 75.4\%  & 0.841                \\
AdaboostM1          & \textbf{77.2\%}   & \textbf{0.843}                \\ \hline
\end{tabular}
}
%\vspace{-5pt}
\end{table}

\begin{figure*}
\centering
\subfigure[Under 2 classes]
{
    \centering
    \includegraphics[width=0.475\textwidth]{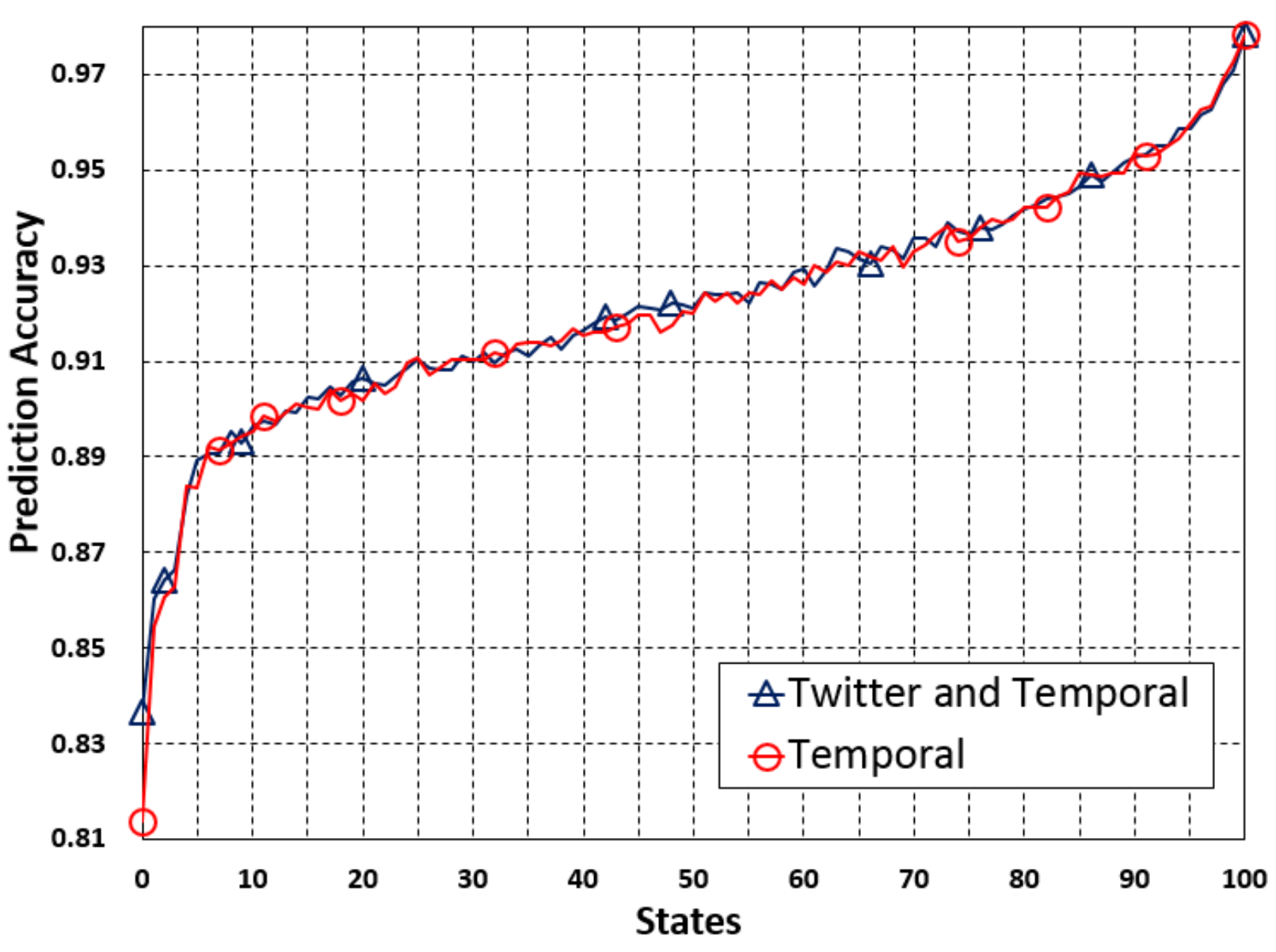}
    %\caption{project successful and the Moving Average with interval 5.}
    \label{fig:TEM-range2}
}
%\hspace{0.5cm}
\subfigure[Under 3 classes]
{
    \centering
    \includegraphics[width=0.475\textwidth]{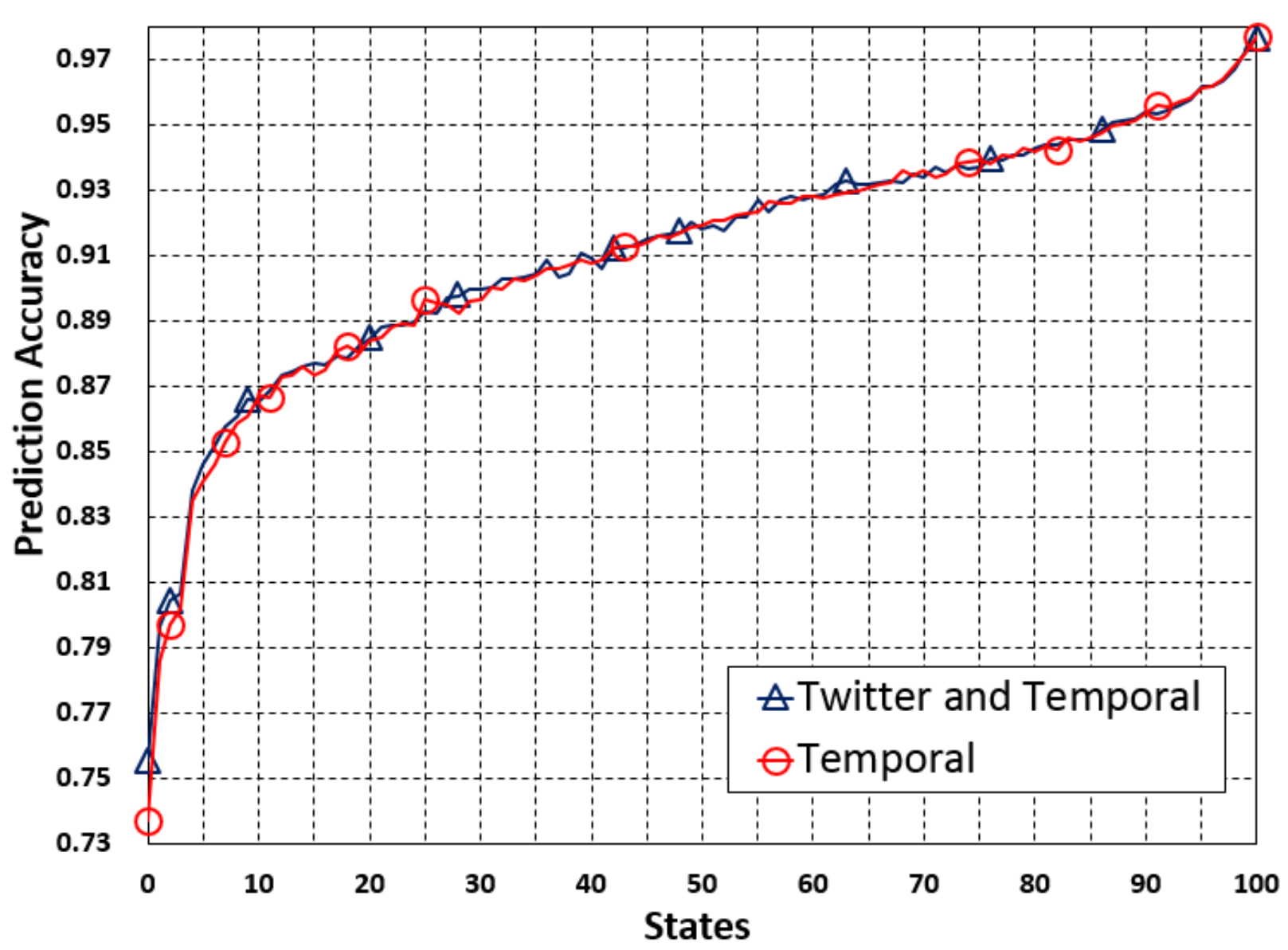}
    %\caption{User joined date and project launched date and the Moving Averages with interval 5.}
    \label{fig:TEM-range3}
}
\caption{Pledged money range prediction rate of predictors based on Kickstarter static and temporal features with/without Twitter features under 2 and 3 classes.}
\label{fig:TEM-Range}
%\vspace{-10pt}
\end{figure*}

\smallskip
\noindent\textbf{Using KS Static + Temporal + Twitter dataset.} What if we add temporal features? Can we find a sweet spot where we can reach to a high accuracy in a short period? To answer these questions, we used KS Static + Temporal + Twitter dataset. Again, each project duration was converted to 100 states (time slots). Figure~\ref{fig:TEM-Range} shows how accuracy of predictors has been changed over time under 2 classes and 3 classes. Prediction accuracy of AdaboostM1 classifiers with all features (project features + user features + temporal features + Twitter features) has been sharply increased until 5th state in 2 classes and 10th state in 3 classes. The classifiers reached to 90\% accuracy in 15th state under 2 classes, and in 31st state under 3 classes.

What if we do not use Twitter features? In both 2 and 3 classes, adding Twitter features slightly increased prediction accuracy until 3rd state in 2 classes, and 9th state in 3 classes compared with predictors without Twitter features.

In summary, our proposed predictive models predicted a project's expected pledged money range with a high accuracy in 2 classes and 3 classes. Adding Twitter and Kickstarter temporal features increased a prediction accuracy even higher than only using Kickstarter static features. Our experimental results confirmed that predicting a project's expected pledged money in advance is possible.

%[KYUMIN: revise it]
\section{Project Creators' Reactions after Projects Failed}
In the previous sections, we found that predicting whether a project will be successful and how much (what range of) fundraising money a project will get. Next, we analyze how project creators behaved after their projects failed. Did they give up and no longer create projects? Or did they continue to create projects? If they continued creating projects with the same idea of the failed projects, what changes did they make in order to make the projects successful.

%Do most creators make one project? How many creators, who totally failed in all previous projects (successful rate = 0), (a) continue creating new project, (b) stop creating new project but still back for other projects they feel interesting or (c) stop creating and backing? What creators do to improve their successful rate over time?

First of all, we analyzed how many projects each user created in Kickstarter as shown in Table~\ref{table:ProjectCreatorDistribution}. 89.74\% (118,718) users created only 1 project while 7.97\% users created 2 projects and 2.29\% users created at least 3 projects. Among the 89.74\% creators, who created only 1 project, 44.15\% project creators successfully reached project goals (i.e., fundraising goals) while 55.85\% project creators failed in reaching project goals. It may mean that the 55.85\% (66,304) project creators among the one-time project creators gave up their project idea, and no longer created new projects.

\begin{table}
	\tbl{Distribution of projects by creators. \label{table:ProjectCreatorDistribution}}
	{
		\centering
		\small
		\begin{tabular}{|c|c|c|}
			\hline
			\# created project	&	\# creators	&	Percentage (\%)  \\ \hline
			1					& 		118,718	&	89.74	\\
			2 					& 		10,546 	& 	7.97	\\
			3 					& 		1,959 	& 	1.48		\\
			4			 		& 		546 	& 	0.41		\\
			5			 		& 		235 	& 	0.18		\\
			${>}$ 5			 	& 		282 	& 	0.21		\\
			\hline
		\end{tabular}
	}
	%\vspace{-5pt}
\end{table}

A follow-up question is ``when a project failed, what properties of the project did project creators change to make the project successful?'' Did they lower project goal? or Did they add more reward types? or Did they add more detailed information into the project description? Before answering these questions, we assume that once a certain project is successful, the project creator will no longer improve or relaunch it. But if a project failed, the project creator may (i) want to improve and relaunch it, (ii) create a project with a completely new idea, or (iii) no longer create any other project. In this study we focus on the first (i) case because we aim to understand what properties of the previously failed project the project creators changed to make it (of the same idea with the previous project) successful.

A challenge in the study was to extract two consecutive projects based on the \emph{same project idea} in chronological order. We assumed that if two consecutive projects created by the same creator were based on the same idea, their project descriptions should be similar. Based on this assumption, we examined 22,320 projects created by 9,166 distinct creators, each of whom created at least 2 projects and had at least one failed project. Then we built Vector Space Model for 22,320 projects so that each project was represented by a TF-IDF based vector \cite{Manning:2008}. We extracted each pair of two consecutive projects created by the same user from the 22,320 projects and measured the cosine (description) similarity of the pair.

Specifically, given two projects $P_i$ and $P_j$ represented by two vectors $V_i$ and $V_j$ respectively, cosine (description) similarity was calculated as follows:
\[
	sim(P_i,P_j) = cos(V_i, V_j) = \frac{\sum_{k=1}^{|D|}v_{ik}v_{jk}}{\sqrt{\sum_{k=1}^{|D|}{v_{ik}^2}}\sqrt{\sum_{k=1}^{|D|}{v_{jk}^2}}}
\]
where, $|D|$ is the total number of unique terms in Vector Space Model, $v_{ik}$ and $v_{jk}$ are TF-IDF values at $k^{th}$ dimension of $V_i$ and $V_j$, respectively.

If a pair's cosine similarity was equal to or greater than a threshold $\lambda$, we would consider the pair as similar projects based on the same project idea.

%\begin{figure}[!t]
%	\centering
%	\begin{floatrow}
%		\ffigbox[\FBwidth]
%		{
%			\caption{Number of pairs of similar projects in which the first one fail and the second one succeed}
%			\label{fig:FailToSuccessPair}
%		}
%		{
%			\includegraphics[width=0.465\textwidth]{fig-files-pdf/FailSuccessSimPair.pdf}
%		}
%		\ffigbox[\FBwidth]
%		{	
%			\caption{Number of pairs of similar projects in which both projects are fail}
%			\label{fig:FailToFailPair}
%		}
%		{
%			\includegraphics[width=0.465\textwidth]{fig-files-pdf/FailFailSimPair.pdf}
%		}	
%	\end{floatrow}
%\end{figure}

\begin{figure}%[!ht]
	\centering
	\subfigure[Number of failed-to-successful project pairs.] % caption for subfigure a
	{
		\label{fig:FailToSuccessPair}
		\includegraphics[width=0.475\textwidth]{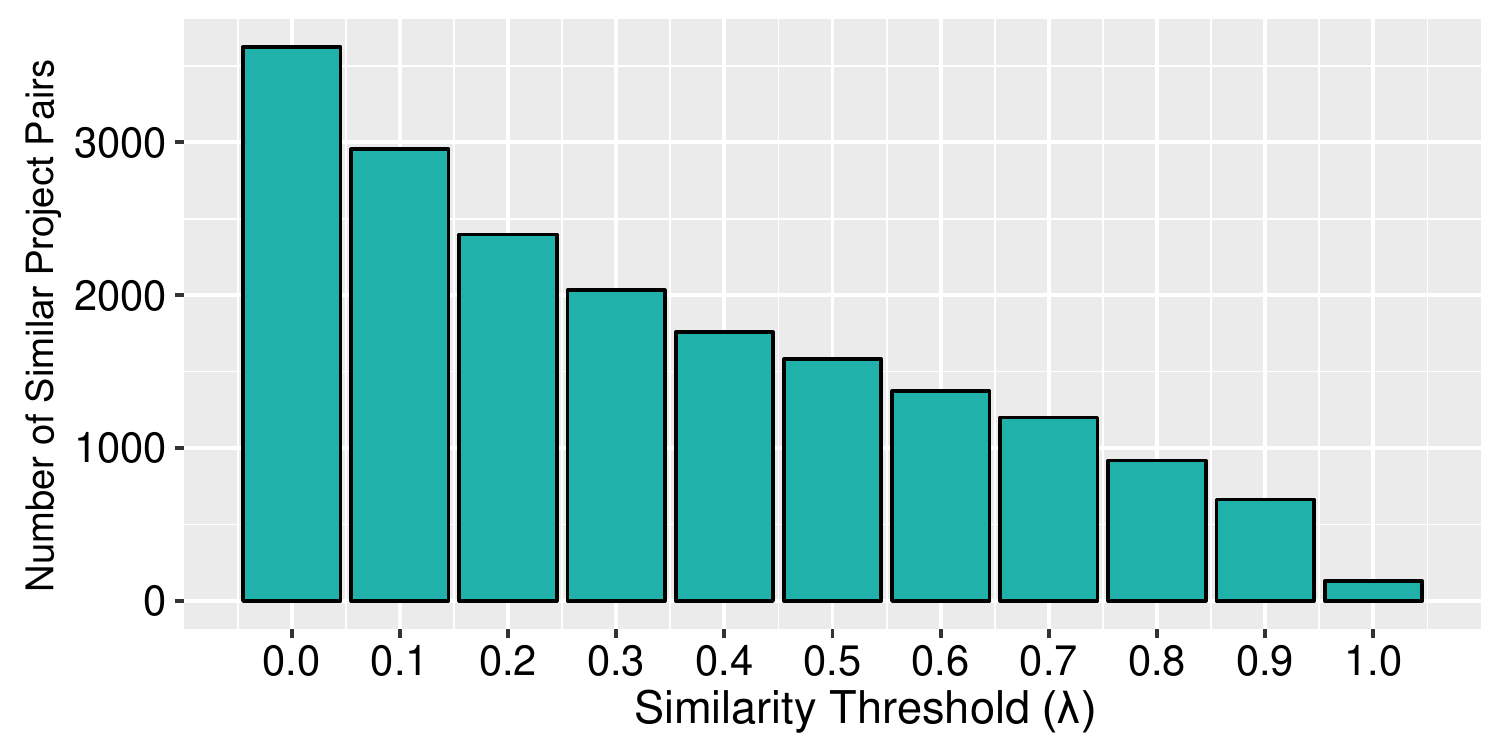}
	}
	%\hspace{0.01cm}
	\subfigure[Number of failed-to-failed project pairs.] % caption for subfigure b
	{
		\label{fig:FailToFailPair}
		\includegraphics[width=0.475\linewidth]{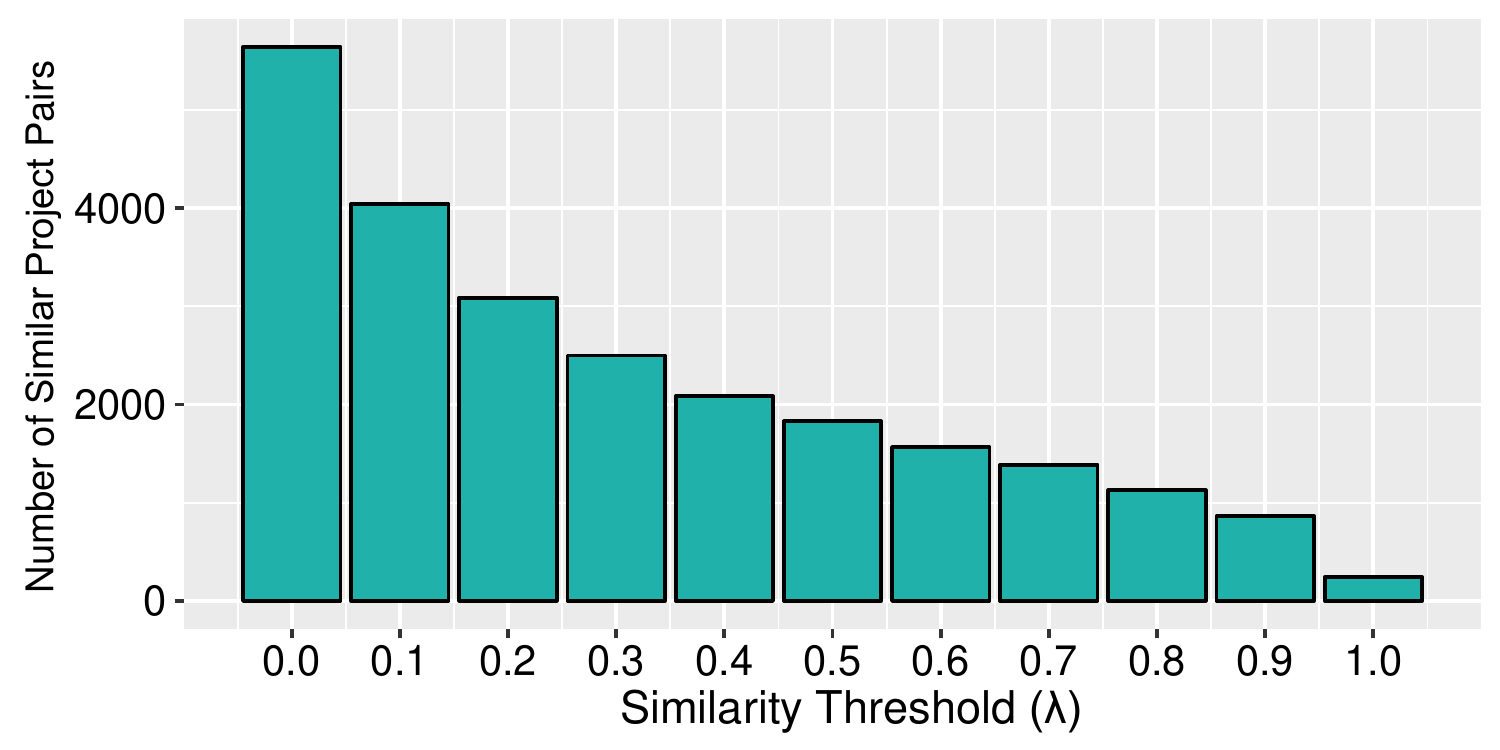}
	}
\caption{Number of similar project pairs in failed-to-successful case and failed-to-failed case.}
\label{fig:pairs}
%\vspace{-10pt}
\end{figure}

%If we set a similarity threshold  for the cosine similarity, we can discover all pairs of similar projects. For each creator, we first extract all projects that he created and sort them by ascending order of timestamp. Each of his projects is mapped to another project with highest similarity score. For example, a creator C created 4 projects: P1, P2, P3, P4. Based on cosine similarity, we discover that P2 is the most similar project of P1 so P1 and P2 form a pair and neither of them are used to map to P3 nor P4.

An up-coming question is what would be a good $\lambda$? To answer this question, first we plotted Figure~\ref{fig:pairs} which shows how the number of pairs of failed-to-failed projects and the number of pairs of failed-to-successful projects were changed as we changed $\lambda$ from 0 to 1 by increasing 0.1. The number of similar project pairs had decreased as we increased $\lambda$. Interestingly, we observed that there were 131 pairs and 242 pairs of projects without changing any word in their project descriptions (i.e., similar score = 1) in Figure~\ref{fig:FailToSuccessPair} and Figure~\ref{fig:FailToFailPair}, respectively. It means some project creators did not change project description of the latter project compared with the former project, but it was successful in 131 cases.

Then, we manually analyzed sample pairs to see what threshold would be the most appropriate to find similar project pairs. Based on the manual investigation, we decided $\lambda$ as 0.8. With the threshold ($\lambda$=0.8), we found 918 failed-to-successful project pairs called \textit{group I} and 1,127 failed-to-failed project pairs called \textit{group II}. By comparing projects in each pair in the two groups, we noticed that overall project creators changed 13 properties: duration, goal, number of images, number of videos, number of FAQs, number of updates, number of rewards, number of sentences in reward description, smog grade of reward, number of sentences in project description, smog grade of project description, number of sentences in project creator's biography, and smog grade of project creator's biography. We measured how much each property was changed by $\frac{(P_{ik} - P_{jk})*100}{P_{ik}}$ where $P_{ik}$ is the former project's \emph{k}th property value and $P_{jk}$ is the latter property's \emph{k}th property value.

Table~\ref{table:ProjectEnhancement} shows the average change rate of failed-to-successful project pairs and failed-to-failed project pairs. A positive change rate means that project creators increased the property value of the latter project compared with the former project. To measure which property had significant difference, we computed one-tailed p-value of two-sample t-test for difference between the means of the two groups. In particular, the mean of project goal's change rate in \textit{group I} was -59.62\%, which was approximately four times decrement compared to \textit{group II} which had -16.39\% change rate. In other words, project creators in \emph{group I} lowered project goal much more than project creators in \emph{group II}. The mean of change rate of the number of updates in \textit{group I} was +118\% while project creators in \textit{group II} made -38.41\% change. It indicates that project creators in \textit{group I} increased the number of updates significantly, while project creators in \textit{group II} decreased the number of updates. Interestingly, decreasing a project duration was helpful to make projects successful. Overall, reducing the duration and goal as well as posting more images, videos and updates are a smart way to make previously failed projects successful.

\begin{figure} %[!ht]
\centering
%\subfigure[Duration] % caption for subfigure a
%{
%    \label{fig:DurantionChange}
%    \includegraphics[width=0.3\textwidth]{more-pdf/Duration-change.pdf}
%}
%\hspace{0.01cm}
\subfigure[Goal] % caption for subfigure b
{
    \label{fig:GoalChange}
    \includegraphics[width=0.48\textwidth]{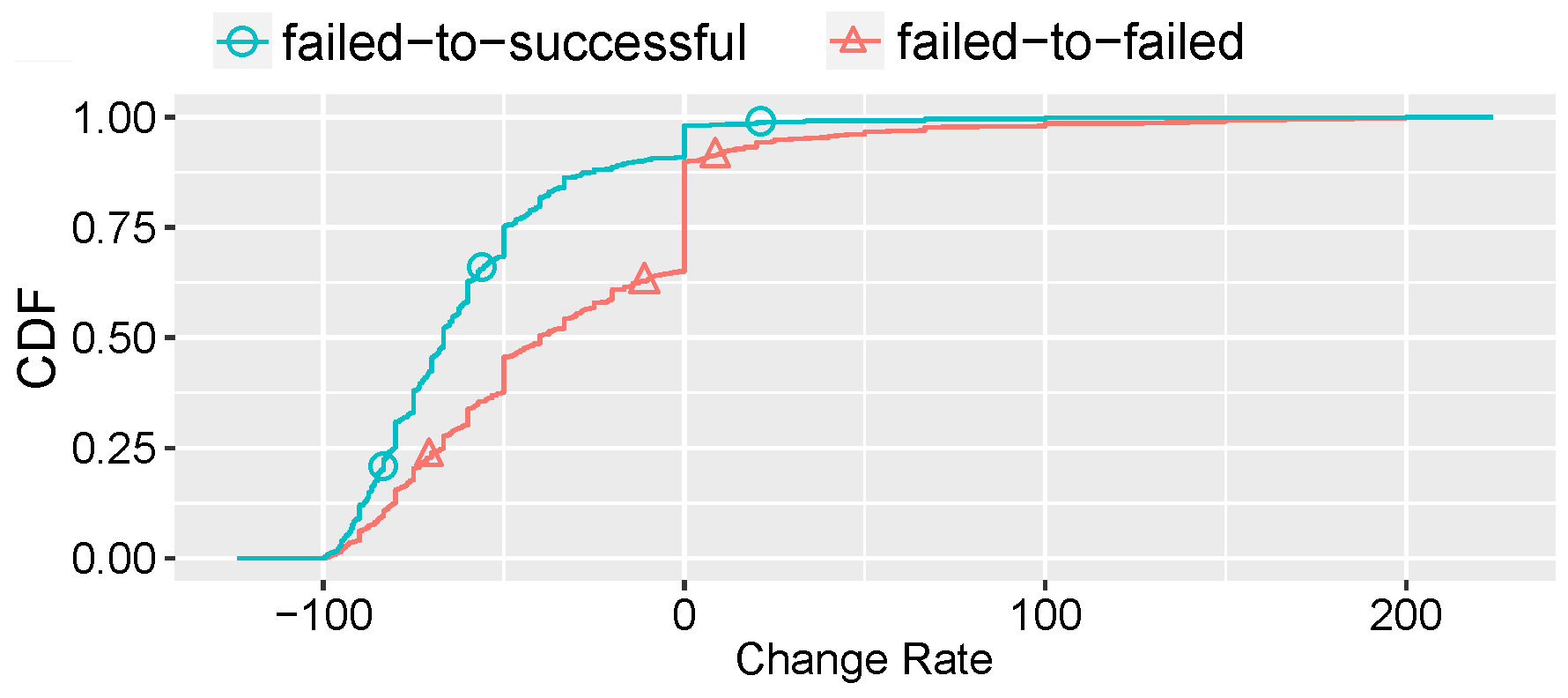}
}
%\hspace{0.01cm}
%\subfigure[Number of Images] % caption for subfigure b
%{
%    \label{fig:NumberOfImageChange}
%    \includegraphics[width=0.3\textwidth]{more-pdf/NumberOfImage-change.pdf}
%}
%\hspace{0.01cm}
%\subfigure[Number of Rewards] % caption for subfigure b
%{
%    \label{fig:NumberOfRewardChange}
%    \includegraphics[width=0.3\textwidth]{more-pdf/NumberOfReward-change.pdf}
%}
%\subfigure[Number of Reward Sentences] % caption for subfigure b
%{
%    \label{fig:NumberofRewardSentenceChange}
%    \includegraphics[width=0.3\textwidth]{more-pdf/NumberofRewardSentence-change.pdf}
%}
%\hspace{0.01cm}
\subfigure[Number of updates] % caption for subfigure b
{
    \label{fig:NumberOfUpdateChange}
    \includegraphics[width=0.48\textwidth]{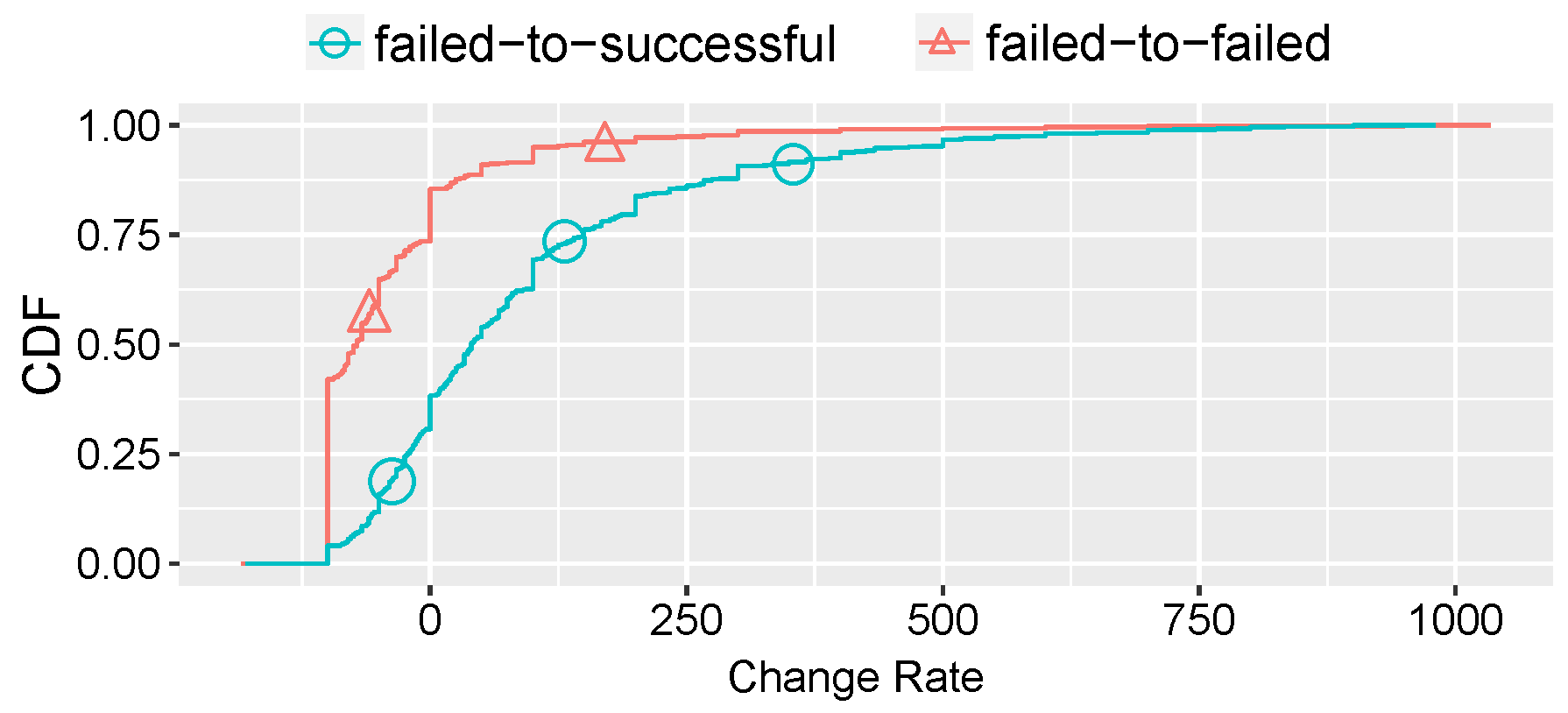}
}
\caption{CDFs of change rates of goal and number of updates in similar project pairs.}
\label{fig:ProjectFeaturesChange} % caption for the whole figure
%\vspace{-10pt}
\end{figure}

Since the number of updates and project goal were the most significant properties, we further analyzed CDFs of change rates of the two properties -- project goal and number of updates -- in the two groups as shown in Figure~\ref{fig:ProjectFeaturesChange}. 88\% project creators in \textit{group I} lowered project goal while 63\% project creators in \emph{group II} lowered project goal. About 62\% project creators in \textit{group I} increased posting the number of updates while only 15\% project creators in \textit{group II} increased posting the number of updates.

\begin{table}%[h]
    \centering \tbl{Average change rate of 13 properties in failed-to-successful project pairs and failed-to-failed project pairs. $***$, $**$, $*$ and ns indicate $p < 10^{-13}$, $p < 10^{-4}$, $p < 0.05$ and \emph{not significant}, respectively.}{
		\small
		\begin{tabular}{|l|>{\centering\arraybackslash}p{3.8cm}|>{\centering\arraybackslash}p{3.8cm}|c|}
			\hline
			Property & Avg. change rate of failed-to-successful pairs \emph{Group I} & Avg. change rate of failed-to-failed pairs \emph{Group II} & p-value \\ \hline
			\textbf{Duration} 		  			& -6.15\%				& +23.03\% 			& **	 \\
			\textbf{Goal}			  	& \textbf{-59.62\%}	& \textbf{-16.39\%} 	& ***	 \\
			\textbf{\#images}			  		& +14.25\%			& +1.91\%				& *		 \\
			\textbf{\#video}					  	& +6.40\%				& -3.22\%				& **	 \\
			\#FAQs					  	& -34.69\%			& -47.47\% 			& ns	 \\
			\#reward		  			& -0.26\%				& +2.36\%				& ns	 \\
			\textbf{\#updates}		  	& \textbf{+118.00\%}	& \textbf{-38.41\%}	& ***	 \\
			smog\_reward			  	& +1.70\%				& +2.78\% 			& ns	 \\
			\#reward\_sentence			& +22.24\%			& +13.73\%			& ns	 \\
			\#main\_sentence		  	& -0.40\%				& -0.27\% 			& ns	 \\
			smog\_main				  	& +7.26\%				& +5.17\% 			& ns	 \\
			\#bio\_sentence			  	& 0\%				& 0\%				& ns	 \\
			smog\_bio				  	& 0\%				& 0\%				& ns	 \\
			\hline
		\end{tabular}
	}
	%\vspace{-5pt}
    \label{table:ProjectEnhancement}
\end{table}

\section{Clustering Successful Projects and Analyzing the Clusters}
In this section, we aim to (i) cluster successful projects based on a time series of normalized daily pledged money, (ii) analyze what kind of clusters we find and how the clusters are different from each other, and understand (iii) how external activities affected projects' temporal patterns. %By clustering successful projects and analyzing the clusters, we can provide intelligent search and discovery for new and existing projects which have similar temporal patterns.

\subsection{Preprocessing Data}
Out of 74,053 projects containing temporal data presented in Table~\ref{table:dataset}, we selected successful projects each of which had a project goal equal to or greater than \$100 since it is less interesting to find patterns from projects whose goal is less than \$100, considering them as noisy data. Finally, the number of the selected projects was 30,333. Since each project has different duration (e.g., 30 days or 60 days), first, we converted each project duration to 20 states (time slots). Then, in each state, we measured obtained pledged money during each state. We created 20 temporal/time buckets and inserted each project's pledged money during each state to each bucket (e.g., the 1st bucket contains each project's pledged money obtained during the first state -- first 5\% duration in this context). To make sure which project got relatively higher or lower pledged money in each bucket, first we measured the mean ($\mu$) and standard deviation ($\sigma$) of pledged money of 30,333 projects in each bucket. Then, we normalized pledged money ($pm_i$) of each project in the \emph{i}th bucket (i.e., pledged money obtained during the \emph{i}th state) as follows:

\[
\bar{pm}_i = \frac{pm_i - \mu_i}{\sigma_i}
\]
where ${\mu_i}$ and ${\sigma_i}$ are the mean and standard deviation of pledged money of the successful projects in \emph{i}th bucket.

After running the normalization in each bucket for the projects, we had a time series of \emph{relative pledged money} for each project, and used these time series in the following subsections.

\subsection{Clustering Approach}
To identify clusters of 30,333 projects, we applied Gaussian Mixture Model (GMM) based clustering algorithm. GMM based clustering approach has been widely used by other researchers in other domains such as clustering experts in a question-answering community \cite{pal2012evolution} and image processing \cite{zivkovic2004improved,permuter2006study}.

We formally define our clustering problem as follows: Given vectors $X=\lbrace{x_1}, {x_2}, ..., {x_N}\rbrace$ of \emph{N} independent projects, where ${x_i}$ represents a time series vector of relative pledged money in \emph{i}th project, we applied GMM based clustering algorithm to find \emph{K} clusters amongst observed \emph{N} time series in \emph{X}.

By using GMM, the log likelihood of the observed \emph{N} time series is written as follows:
%Since we did normalize 100 daily pledged fund of each project into 20 5-days buckets, each vector ${x_i}$ has dimension \textit{D} = 20
\[
lnP(X\mid\pi,\mu,\Sigma) = \sum_{i=1}^{N}ln \bigg\{ \sum_{k=1}^{K}\pi_k \mathcal{N}(x_i\mid\mu_k, \Sigma_k) \bigg\}
\]
, where the parameter $\{{\pi_k}\}$ is the mixing coefficients of a cluster \emph{k} and must satisfy two conditions: ${0\leq\pi_k\leq1}$ and ${\sum_{k=1}^{K}\pi_k} = 1$. $\mu_k$ and $\Sigma_k$ are the mean and covariance matrix of the cluster \emph{k}, respectively. ${\mathcal{N}(x_i\mid\mu_k, \Sigma_k) }$ is the multivariate Gaussian distribution of cluster \emph{k}, defined as follows:
\[
\mathcal{N}(x_i\mid\mu_k, \Sigma_k) = \frac{1}{(2\pi)^{D/2}}\frac{1}{\mid\Sigma_k\mid^{1/2}} exp \bigg\{ -\frac{1}{2} (x_i -\mu_k)^T \Sigma_k^{-1} (x_i -\mu_k)  \bigg\}
\]
We used EM algorithm to maximize the log likelihood function with regard to parameters including means ${\mu_k}$, covariance  ${\Sigma_k}$ and the mixing coefficient ${\pi_k}$. We first initialized the values of these parameters. Then in Expectation step, the responsibilities ${\gamma_k(x_i)}$ of the $k^{th}$ component of observation $x_i$ was calculated by the current parameter values with regard to Bayesian theorem as follows:
\[
\gamma_k(x_i) = p(k|x_i)
			  = \frac{p(x_i)p(x_i|k)}{\sum_{l=1}^{K}{p(l)p(x_i|l)}}
			  = \frac{\pi_k \mathcal{N}(x_i\mid\mu_k, \Sigma_k) }{ \sum_{j=1}^{K} \pi_j   \mathcal{N}(x_i\mid\mu_j, \Sigma_j)}
\]
In Maximization step, parameters ${\mu_k}$, ${\Sigma_k}$ and ${\pi_k}$ were re-estimated by using the current responsibilities as follows:
\[
	\mu_{k}^{new} = \frac{1}{ \sum_{i=1}^{N} \gamma_k(x_i) }  \sum_{n=1}^{N}\gamma_k(x_i)x_i	
\]
\[
	\Sigma_{k}^{new} = \frac{1}{ \sum_{i=1}^{N} \gamma_k(x_i) }  \sum_{n=1}^{N}\gamma_k(x_i) (x_i-\mu_k^{new}) (x_i -\mu_k^{new})^T
\]
\[
	\pi_k^{new} = \frac{\sum_{i=1}^{N} \gamma_k(x_i)}{N}
\]
Then, the log likelihood was evaluated. The EM algorithm was stopped when the convergence condition of log likelihood was satisfied or the number of iterations exceeded a pre-defined value.

To estimate the optimal number of clusters inputting in GMM, we used the Bayesian Information Criteria (BIC). In statistics, BIC is a criterion based on the likelihood function for model selection among a finite set of models. The model with the lowest BIC value is the best one among the models. In our study, a model with the lowest BIC value indicates that the number of clusters \emph{K} in the model is the optimal number, returning the most meaningful clusters. Let ${\widehat{L}}$ as the maximum value of the likelihood function of the model, the value of BIC is calculated as following:
\[
	BIC(K) = -2ln\widehat{L} + K\ln{N}
\]

\begin{figure}%[!ht]
\centering
\includegraphics[width=0.465\linewidth]{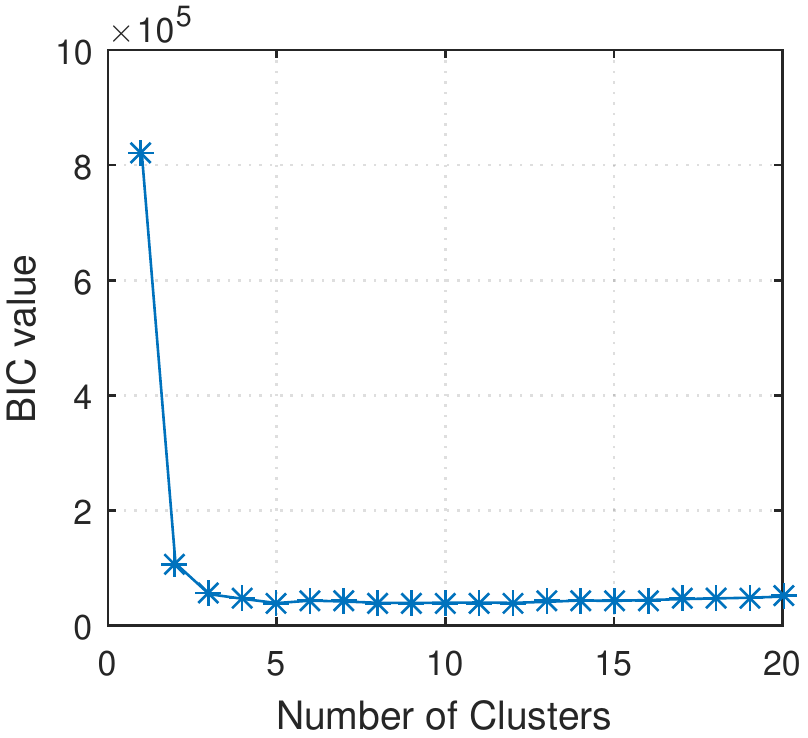}
\caption{A BIC curve of the 30,333 successful projects.}
\label{fig:BIC-values}
%\vspace{-10pt}
\end{figure}

%\begin{figure*}[t]
%\centering
%\subfigure[Successful projects]
%{
%    \centering
%    \includegraphics[width=0.465\textwidth]{fig-files-pdf/BIC-Successful.pdf}
%    \label{fig:BIC-suc}
%}
%\hspace{0.1cm}
%\subfigure[Failed projects]
%{
%    \centering
%    \includegraphics[width=0.465\textwidth]{fig-files-pdf/BIC-Fail.pdf}
%    \label{fig:BIC-fail}
%}
%\caption{Plots of BIC curves}
%\label{fig:BIC-values}
%\vspace{-10pt}
%\end{figure*}

%\smallskip

%\noindent\textbf{Number of clustering patterns:} We separately run the GMM based clustering algorithm to find interesting patterns existed among successful and failed projects. Figure~\ref{} shows the BIC curve as a function of K for different runs of GMM algorithm (K = [1,..,20]) in the successful and failed projects. We pick K = 5 as the number of clusters in both these two groups since it minimizes the BIC criteria in all 20 runs of the GMM.

%We then run GMM with 5 clusters as input and aggregate the normalized mean series of projects in each cluster to compute the cluster aggregate series. Figure \ref{} shows the patterns of normalized pledged fund for different clusters found by GMM in successful projects and failed projects, respectively. For each group, we discover some interesting patterns as described below:

%\subsection{Popularity of projects over time}
\subsection{Analysis of Clusters}
To find the optimal number of clusters, we ran the GMM based clustering algorithm in a range of $K = {1 \sim 20}$ by increasing 1 in each time, and got a BIC value in each case. Figure~\ref{fig:BIC-values} depicts a BIC curve showing how a BIC value was changed as we increased \emph{K} by 1 in each time. Finally, $K$ = 5 returned the smallest BIC value and returned the optimal 5 clusters.

\begin{figure}%[!ht]
\centering
\includegraphics[width=0.465\linewidth]{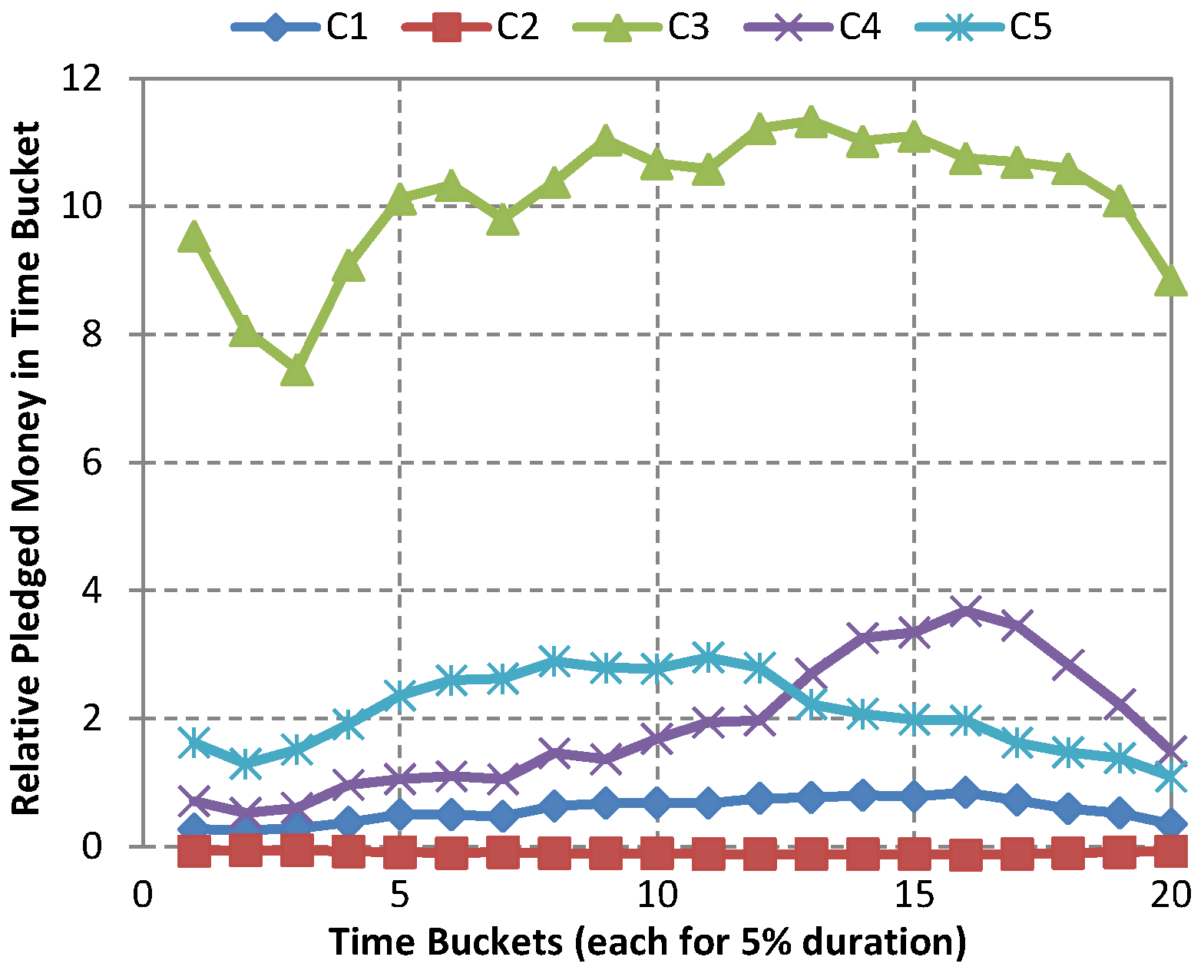}
\caption{Evolutional patterns of five clusters.}
\label{fig:pattern-suc}
%\vspace{-10pt}
\end{figure}

To understand how each cluster had different temporal patterns, we measured the mean of relative pledged money in each bucket of projects in each cluster. Then, we drew a line of the means for each of the five clusters as shown in Figure~\ref{fig:pattern-suc}.

\begin{itemize}
  \item Projects in a cluster C2 received almost same amount of relative pledged money over time.
  \item Projects in a cluster C3 received the largest amount of pledged money over time compared with projects in the other four clusters. In the beginning, relative pledged money went down until the 3rd time bucket, went up until the 13th time bucket with some fluctuation, and then gradually went down. Why did this evolutional pattern happen? We conjecture that the news of initial popularity was propagated to other users, some of whom eventually backed up the projects, increasing daily/relative pledged money. It is a typical evolutional pattern of the most popular projects like the Coolest Cooler \cite{cooler} and the Pono Music \cite{pono}\footnote{The Coolest Cooler project received \$13,285,226, and the Pono Music project received \$6,225,354.}.
  \item A cluster C4 had the most interesting pattern. The initial popularity (pledged money) was low, but the pledged money gradually increased until the 16th time bucket with sharp increments between 12th and 14th time buckets. A cluster C1 (less interesting cluster) had a similar pattern with C4, but overall increments were much lower than C4.
  \item A cluster C5 had also an interesting pattern which was gradually going up during the first half duration and going down during the other half duration.
  \end{itemize}

\begin{table}[t]
\tbl{Number of projects, average project goal and average pledged money in each cluster.
\label{table:distribution-success}}
{
\centering
\small
\begin{tabular}{|c|r|r|r|}
\hline
Cluster     & $|$projects$|$ & Avg. goal & Avg. pledged money \\ \hline
C1          & 1,563               &  \$41,542  &    \$95,429 					    \\
C2          & 28,209              &   \$6,334  &     \$9,306                      \\
C3          & 97                 & \$273,222  & \$1,487,672                      \\
C4          & 186                &  \$98,253  &  \$227,078                      \\
C5          & 278                &  \$79,354  &   \$284,761                      \\ \hline
\end{tabular}
}
\end{table}

Next, we analyzed how many projects belonged to each cluster, and estimated average project goal and pledged money of projects in each cluster. Table~\ref{table:distribution-success} shows the number of projects, and corresponding average project goal and average pledged money. Two largest clusters were C2 and C1 consisting of 28,209 (93\%) and 1,563 (5\%) projects, respectively. These clusters had the lowest goal, and achieved the lowest pledged money compared with the other three clusters. C3 had the highest goal and got the highest pledged money. C4 and C5 had next highest goal and got next highest pledged money. Overall, each of the top 2\% successful projects (including C3, C4 and C5) on average received more than 200K pledged money. It means that there were a lot of successful projects with low goal and low pledged money, while there existed a small portion of projects (2\%) with high goal and high pledged money, resulting in unequal distribution of pledged money across successful projects in a crowdfunding platform, Kickstarter.

\begin{table}%[h]
\tbl{Average percent of duration reaching a goal in each cluster.
\label{table:TimeforGoal}}
{
\centering
\small
\begin{tabular}{|c|r|c|}
\hline
Cluster     & Avg. goal & Avg. percent of duration reaching a goal \\ \hline
C1          & \$41,542   & 55\%     				       \\
C2          & \$6,334    & 66\%                            \\
C3          & \$273,222  & 17\%                            \\
C4          & \$98,253   & 58\%                           \\
C5          & \$79,354   & 26\%                            \\ \hline
\end{tabular}
}
\end{table}

\begin{table}%[h]
\tbl{Average property values in each cluster.
\label{table:SUC-proj-features}}
{
\centering
\small
\begin{tabular}{|c|c|rrrrrr|}
\hline
\multirow{2}{*}{Cluster} & \multicolumn{7}{c}{Average}    \\ \cline{2-8}
                           & Pl. Money & $|$Images$|$ & $|$Videos$|$ & $|$FAQs$|$ & $|$Rewards$|$ & $|$Updates$|$ & $|$Comments$|$ \\  \hline
C1         & 85,429            & 18.36        & 2.03         & 3.38        & 15.51         & 19.94         & 405.70           \\
C2         & 9,306            & 6.59         & 1.28         & 0.72       & 10.07         & 9.14          & 26.89             \\
C3         & 1,487,672            & 34.44        & 2.51         & 12.71      & 18.20         & 41.80         & 16,712.34          \\
C4         & 227,077            & 23.74        & 2.52         & 5.50       & 18.89         & 27.28         & 1509.78           \\
C5         & 284,761            & 22.24         & 2.20         & 7.66       & 14.57         & 23.94         & 1233.47           \\ \hline
\end{tabular}
}
\end{table}

Up-coming questions are ``When did projects in each cluster reach their goal? Did they reach in almost similar time (e.g., the first 30\% duration)?''. To answer these questions, we analyzed accumulated daily pledged money to see when they reached the goal. Table~\ref{table:TimeforGoal} presents the analytical results. All the successful projects reached their goal before 67\% duration. Projects in cluster C3 (with the highest goal and pledged fund) reached their goal very fast, only in 17\% duration. Projects in C5 reached their goal faster than projects in C4, but total pledged money was less than C4 in the end of the fundraising campaigns. Interestingly, projects in C1, which had similar (but less popular) temporal pattern with C4 in Figure~\ref{fig:pattern-suc}, reached their goal in similar time (55\%) even though their goal was lower than C4. C2 with the lowest goal took the longest duration to reach the goal.

Next, we further analyzed the five clusters to understand how other properties were associated with pledged money across the five clusters. In particular, we focused on properties such as number of updates, number of images, number of videos, number of FAQs, number of rewards, number of updates and number of comments. Table~\ref{table:SUC-proj-features} shows the average value of the properties in each cluster. We clearly observed that projects in C3 had the largest values in all the properties except the number of videos (still almost similar with the largest value in C4). Project creators in C3 spent more time to create their project descriptions by adding more images, videos and reward types. During a fundraising period, they actively added more updates, FAQs and received more comments from backers. Mostly, these phenomena applied to the other clusters.

Finally, we focused on C4 and C5 which had interesting evolutional patterns as shown in Figure~\ref{fig:pattern-suc}. Specifically, projects in C4 were initially not popular, but later became popular with a sharp increment in terms of relative pledged money in each time bucket, while projects in C5 were initially popular and then became less popular or relative pledged money in each time bucket decreased. To understand the phenomenon, we investigated how external promotional activities in C4 and C5 were different.

\begin{figure}[t]
	\centerline{
		\includegraphics[width=0.5\textwidth]{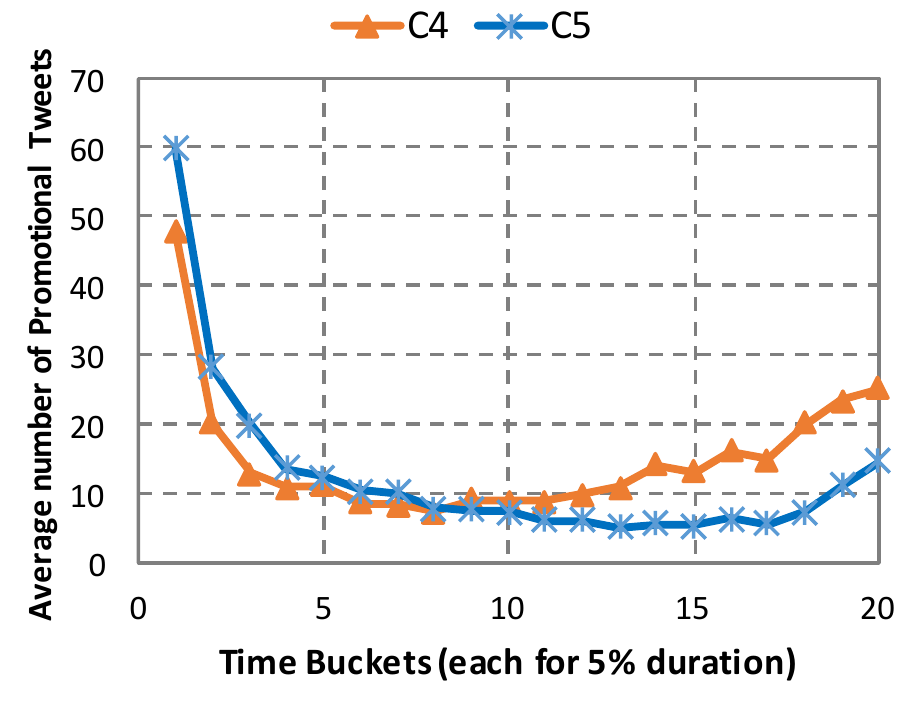}
	}
	\caption{Average number of promotional tweets posted during each time bucket in C4 and C5.}
	\label{fig:PromotionPattern}
	%\vspace{-10pt}
\end{figure}

To conduct this study, first we collected promotion-related tweets for each project in C4 and C5 from Twitter by searching each Kickstarter project URL. These tweets were posted by project creators, their friends and backers. Then, we computed the average number of promotion tweets during each time bucket in each cluster. Figure~\ref{fig:PromotionPattern} shows how the number of promotion tweets was changed over time. Interestingly, in the first 8 time buckets, the number of promotion tweets in C5 were higher than the number of promotion tweets in C4. Since then, the situation was reversed -- there were more promotion tweets in C4 than C5. Interestingly, the temporal promotional activities were similar with the evolutional patterns of pledged money in C4 and C5 shown in Figure~\ref{fig:pattern-suc}. Note that it took time for these promotional activities to take effect in terms of relative pledged money in each time bucket. Based on this study, we conclude that promotional activities on social media played an important role for increasing relative pledged money over time.

\section{Discussion}
In Sections~\ref{sec:features},~\ref{sec:success} and~\ref{sec:range}, we described our proposed approaches with a list of feature, and showed experimental results. In this section, we discuss other features that we tried to use but finally excluded because of degrading performance of our predictive models.

\subsection{N-gram Features}
In the literature, researcher have generated and used n-gram features from texts such as web pages, blogs and short text messages toward building models in various domains like text categorization \cite{text:categorization}, machine translation \cite{Marioo:2006} and social spam detection \cite{Lee:2010}.

We extracted unigram, bigram and trigram features from Kickstarter project descriptions after lowercasing the project descriptions, and removing stop words. Then, we conducted $\chi^{2}$ feature selection so that we could only keep n-gram features which have positive power distinguishing between successful projects and failed projects. Finally, we added 22,422 n-gram features to our original feature set (i.e., project features, user features, temporal features and Twitter features) described in Section~\ref{sec:features}. Then, we built and tested project success predictors. Unfortunately, adding n-gram features deteriorated performance of project success predictors compared with only using the original feature set described in Section~\ref{sec:features}. The experimental results were the opposite of our expectation because other researchers \cite{Mitra:2014} reported that using n-gram features improved their prediction rate in their own Kickstarter dataset. We conjecture that the researchers used smaller dataset which might give them some improvements. But, given the larger dataset containing all Kickstarter projects, using n-gram features decreased a prediction rate.

\subsection{LIWC Features}
We were also interested in using the Linguistic Inquiry and Word Count (LIWC) dictionary, which is a standard approach for mapping text to psychologically-meaningful categories \cite{james2001linguistic}, to generate linguistic features from a Kickstarter project main description, reward description and project creator's bio description. LIWC-2001 defines 68 different categories, each of which contains several dozens to hundreds of words. Given a project's descriptions, we measured linguistic characteristics in the 68 categories by computing a score of each category based on LIWC dictionary. First we counted the total number of words in the project description (\emph{N}). Next we counted the number of words in the description overlapped with the words in each category \emph{i} on LIWC dictionary ($C_{i}$). Then, we computed a score of a category \emph{i} as $C_{i}/N$. Finally, we added 68 features to the original features described in Section~\ref{sec:features}. Then we built project success predictors and evaluated their performance. Unfortunately, the predictors based on 68 linguistic features and the original features were worse than predictors based on only the original features.

%In general a short summarizing paragraph will do, and under no circumstances should the paragraph simply repeat material from the Abstract or Introduction. In some cases it's possible to now make the original claims more concrete, e.g., by referring to quantitative performance results.

\section{Conclusion}
In this manuscript we have analyzed users and projects in Kickstarter. We found that 46.1\% users were all-time creators and 53.9\% users were active users who not only created their own projects but also backed other projects. We also found that project success rate in each month has been decreasing as new users joined Kickstarter and launched projects without enough preparation and experience. When we analyzed temporal data of our collected projects, we noticed that there were two peaks in the beginning of a project duration and there was the deadline effect, rushing to invest the project as the project was heading to the end of its duration. Then, we proposed 4 types of features toward building predictive models to predict whether a project will be successful and a range of pledged money. We developed the predictive models based on various feature sets. Our experimental results have showed that project success predictors based on only static features achieved 76.4\% accuracy and 0.838 AUC, by adding Twitter features, increased accuracy and AUC by 2.5\% and 3.5\%, respectively. Adding temporal features consistently increased the accuracy. Our pledged money range predictors based on the static features have achieved up to 86.5\% accuracy and 0.901 AUC. Adding Twitter and temporal features increased performance of the predictors further.

We analyzed what reactions project creators made when their projects failed. By identifying similar project pairs, we compared what properties project creators changed in order to make their failed projects successful in the next try. Our t-test revealed that project creators who lowered their project goal by -59.62\% and increased posting the number of updated by +118\% on average made the projects successful. Then, we clustered successful projects based on a time series of relative pledged money, and found 5 clusters. Out of the 5 clusters, we found three interesting clusters: (i) projects in a cluster were the most popular, receiving the highest relative pledged money over time; (ii) relative pledged money of projects in a cluster went up and went down; and (iii) relative pledged money of projects in a cluster had low relative pledged money initially, but went up with a sharp increment. Overall, our work will help project creators organize their projects intelligently, creating better project description and behaving more actively while running fundraising campaigns, and eventually increasing project success rate.

% Acknowledgments
%\begin{acks}
%The authors would like to thank Dr. Maura Turolla of Telecom
%Italia for providing specifications about the application scenario.
%\end{acks}

% Bibliography
\bibliographystyle{ACM-Reference-Format-Journals}
\bibliography{acmsmall-sample-bibfile}
                             % Sample .bib file with references that match those in
                             % the 'Specifications Document (V1.5)' as well containing
                             % 'legacy' bibs and bibs with 'alternate codings'.
                             % Gerry Murray - March 2012

% History dates
%\received{February 2007}{March 2009}{June 2009}

\end{document}